# Mobility and Cost Aware Inference Accelerating Algorithm for Edge Intelligence

Xin Yuan, Ning Li, Kang Wei, Wenchao Xu, Quan Chen, Hao Chen, Song Guo, *Fellow, IEEE*

**Abstract**—The edge intelligence (EI) has been widely applied recently. Splitting the model between device, edge server, and cloud can improve the performance of EI greatly. The model segmentation without user mobility has been investigated deeply by previous works. However, in most use cases of EI, the end devices are mobile. Only a few works have been carried out on this aspect. These works still have many issues, such as ignoring the energy consumption of mobile device, inappropriate network assumption, and low effectiveness on adapting user mobility, etc. Therefore, for addressing the disadvantages of model segmentation and resource allocation in previous works, we propose mobility and cost aware model segmentation and resource allocation algorithm for accelerating the inference at edge (MCSA). Specifically, in the scenario without user mobility, the loop iteration gradient descent (Li-GD) algorithm is provided. When the mobile user has a large model inference task needs to be calculated, it will take the energy consumption of mobile user, the communication and computing resource renting cost, and the inference delay into account to find the optimal model segmentation and resource allocation strategy. In the scenario with user mobility, the mobility aware Li-GD (MLi-GD) algorithm is proposed to calculate the optimal strategy. Then, the properties of the proposed algorithms are investigated, including convergence, complexity, and approximation ratio. The experimental results demonstrate the effectiveness of the proposed algorithms.

**Index Terms**—Edge Intelligence; Model Segmentation; Inference Accelerating; Mobility

———————————————————— ◆ ————————————————————

## 1 INTRODUCTION

The artificial intelligence has been widely used and changed our life greatly, such as metaverse [1-2], automatic driving [2-4], image generation [5], etc. However, since the AI model is always large for achieving high accuracy, the computing resource that needed for these models are huge. Therefore, it is inappropriate to deploy these AI models on the mobile devices, such as mobile phones and vehicles, in which the computing resource is quite limited. For addressing this issue, one possible solution is to divide the large AI model into different parts and offload the resource-intensive model to edge server for reducing resource requirement and latency [6-8].

The model segmentation between end devices, edge server and cloud without user mobility has been investigated deeply by previous works, such as [9-16]. These works find the optimal model segmentation point and early-exist point to minimize inference delay and the allocated resource while maintain high inference accuracy by reforcement learning [9], convex optimization [10-13], heuristic algorithm [14-16], etc. Based on these approaches, many large AI models have been deployed on resource limited end devices successfully to provide high quality intelligent services, like AR/VR, automatic drive, image identification, etc. However, in most scenarios of edge computing and edge intelligence, the end devices are mobile, such as vehicular edge computing [17]. Once the mobile users move out of the coverage of edge server that they belong to, then how to guarantee the qualiy of service (delay, accuracy, energy consumption, etc.) for these users is challenging. Therefore, how to adaptive the model segmentation and resource allocation to user mobility becomes attractive to the researchers.

To the best of our knowledge, until now only the works in [18], [19] and [20] have taken the user mobility into account to improve the performance of model inference. However, they have disadvantages as follows. First, for the mobile users, once they move into the coverage of a new edge server, due to the heterogeneity of edge servers, the mobile users have multiple alternative strategies, such as transmitting inference task back to the original edge server, migrating the partial model that offload to the original edge server to the new one, or recomputing the model segmentation strategy, etc. Then how to find the most optimal strategy from these alternative choices is important to the inference performance. Unfortunately, the works in [18] and [19] focus on frame scheduling for minimizing service interrupt time during handover, the works in [20] focus on the tradeoff between model segmentation and connection reliability in VEC. Second, for the mobile devices, one of the most important factors to consider is the energy consumption, because the energy consumption relates to the lifetime of devices. In most mobility scenarios, the lifetime of mobile devices is more important than inference delay

————————————————
- *F.A. Author is with the National Institute of Standards and Technology, Boulder, CO 80305. E-mail: author@ boulder.nist.gov.*
- *S.B. Author Jr. is with the Department of Physics, Colorado State University, Fort Collins, CO 80523. E-mail: author@colostate.edu.*
- *T.C. Author is with the Electrical Engineering Department, University of Colorado, Boulder, CO 80309. On leave from the National Research Institute for Metals, Tsukuba, Japan E-mail: author@nrim.go.jp.*

***Please provide a complete mailing address for each author, as this is the address the 10 complimentary reprints of your paper will be sent***

*Please note that all acknowledgments should be placed at the end of the paper, before the bibliography (note that corresponding authorship is not noted in affiliation box, but in acknowledgment section).*

and accuracy to some extent, because long lifetime of devices is the fundamental of high QoS [21-23]. Once the mobile devices are out of energy, the QoS is meaningless. Therefore, the energy consumption of mobile devices is critical to edge intelligence and should be considered during model segmentation. However, this has not been investigated by the previous works with or without user mobility. Finally, the previous works assume that one access point (AP) can associate with one edge server and the mobile users can access to the edge server within one hop. However, this assumption is inappropriate in practice. Since in most cases, not all the APs can be arranged an edge server due to the deployment cost. One edge server may serve more than one APs, and the mobile users within the coverage of APs that do not associate with edge servers need to transmit their tasks to the adjacent APs which associate with edge servers by multi-hops manner [24-26]. Thus, when one mobile user moves from the coverage of edge server to another, the optimal task transmission routing to the new edge server will be changed, which will cause the changing of transmission delay. Moreover, considering the heterogeneity of different edge servers (i.e., different computing capability), the original model segmentation and resource allocation strategy may be not optimal in the new edge server anymore, which will cause performance deterioration. However, this is not considered by the previous works. Additionally, in practice, the amount of computing resource in edge sever is always limited, especially when the number of users is large [14], which is also not considered in [18], [19] and [20].

Solving the above issues is challenging. Firstly, finding the optimal model segmentation and resource allocation strategy when considering both energy consumption, resource limitation, and inference delay is not easy. Because the optimal objectives of these three parameters are opsite. For instance, when minimizing the allocated resource of edge server, the size of model that offloading to edge server should be as small as possible. However, this means most part of the model shoule be calculated on device, which will cause high inference delay and energy consumption of end device. Therefore, how to find the optimal tradeoff between these three parameters and optimize the performance of the whole systems is difficult. Secondly, as introduced above, when the mobile user moves into the coverage of a new edge server, they will have multiple alternative choices. However, condidering the multi-user scenario, user mobility, heterogeneity edge devices, and dynamic network condition, finding the most optimal strategy from these alternative strategies to achieve high inference QoS is also challenging.

Based on the above disadvantages of previous works, in this paper, the Mobility and Cost (energy consumption and resource renting cost) aware model Segmentation and resource Allocation algorithm is proposed for accelerating the inference at edge, shorted as MCSA. Specifically, in the scenario without user mobility, when the mobile user has a large model inference task needed to be calculated, it will take the energy consumption of mobile user, the communication and computing resource renting cost, and the inference delay into consideration to find the optimal model segmentation strategy and resource allocation strategy. Since the minimum delay, minimum cost, and minimum energy consumption cannot be satisfied simultaneously, the gradient descent (GD) algorithm is adopted to find the optimal tradeoff between them. Moreover, the loop iteration GD approach (Li-GD) is proposed to reduce the complexity of GD algorithm that caused by the discrete of model segmentation. In the scenario which considering the user mobility, there are three alternative strategies: 1) recalculating the optimal model segmentation and resource allocation strategy when the user moves into the coverage of new edge server, 2) transmitting the inference task back to the original edge server, or 3) migrating the splitted model that on the original edge server to the new one. For finding the optimal solution between these alternative strategies, variable $R$ is introduced into the objective function to represnt the alternative strategies. Then, we propose the mobility aware Li-GD (MLi-GD) algorithm to calculate the optimal model segmentation, resource allocation, and strategy selection under user mobility. Moreover, the properties of the proposed algorithms are also investigated, including convergence, complexity, and approximation ratio.

The contributions of this paper can be summarized as follows.

1) In this paper, for prolonging the lifetime of end devices, and minimizing the resource usage of edge server and the inference delay, both the energy consumption, the resource renting cost, and the inference delay are considered to find the optimal model segmentation and resource allocation strategy in both the scenarios with and without user mobility. The more practically inference delay model, energy consumption model, and resource renting cost model are proposed in this paper. Additionally, since the optimal objectives of these three parameters are opsite, the GD approach is used to find the optimal tradeoff between these three parameters. For reducing the complexity of GD algorithm caused by the discrete of model segmentation, the Li-GD is proposed, which is effective on reducing the complexity of GD algorithm. Moreover, the MLi-GD algorithm is also proposed for the scenario with user mobility.

2) The effect of user mobility on model segmentation and resource allocation is considered, and the assumption is more practically than previous works. First, considering that one edge server can serve more than one APs, the more pratically assumption is provided in this paper. Then, based on this assumption and considering the mobility of end device, for improving the performance of model segmentation and resource allocation, the "model-mule" is introduced into the algorithm, in which the mobile user stores the whole inference model and moves around. When the mobile users move into the coverage of a new edge server, they can calculate the most optimal model segmentation and resource allocation strategy from the alternative strategies more effectively than previous works.

3) In this paper, the properties of the proposed Li-GD algorithm and MLi-GD algorithm are investigated. First, it can be proved that the Li-GD algorithm is convergent, and the convergence time is $K = \frac{\|x^0 - x^*\|_2^2}{2\eta\epsilon}$, the complexity of the Li-GD is $O(X\bar{K}Mx^3\ln^2(x))$. Additionally, it can be demonstrated that the Li-GD algorithm can reduce complexity and convergence time. Then, for the MLi-GD algorithm, it can be verified that the convergence time and the complexity of MLi-GD algorithm are the same as the Li-GD algorithm, only the calculation of the final optimal result is a little more complex than that in the Li-GD algorithm. The approximation ratio of MLi-GD algorithm is $\epsilon$.

The rest of this paper is organized as follows. Section II introduces the related works. The network models and the problems that will be solved in this paper is presented in Section III. The Li-GD algorithm for the scenario without user mobility is proposed in Section IV, and the properties, e.g., the convergence, the complexity, are also investigated in this section. The MLi-GD algorithm for the scenario with user mobility and its properties are presented in Section V. In Section VI, the effectiveness of the proposed MCSA algorithm is demonstrated by simulation. Section VII summarizes the conclusions of this work.

## 2 RELATED WORKS

### 2.1 Model segmentation and resource allocation without user mobility

For the model segmentation and resource allocation without user mobility, the research efforts focus on offloading computation from the resource-constrained mobile to the powerful cloud to reduce inference time [27],[28]. Neurosurgeon [29] explores a computation offloading method for DNNs between the mobile device and the cloud server at layer granularity. In [30], the authors propose MAUI, which is an offloading framework that can determine to execute the functions of a program on edge or cloud. But the MAUI is not explicitly designed for DNN segmentation since the size of communication data between devices and edge/cloud is small. The DDNN is proposed in [31], which is a distributed deep neural network architecture that is distributed across different computing devices. The purpose of DDN is to reduce the communication data size among devices for the given DNN. The authors in [9] propose multi-exit DNN inference acceleration framework based on multi-dimensional optimization. This is the first work that investigates the bottlenecks of executing multi-exit DNNs in edge computing and builds a novel model for inference acceleration with exit selection, model partition, and resource allocation. The authors of [32] and [33] propose feature-based slicing method, which divided a layer into multiple slices for parallel inference. The [32] established a mobile devices' cluster and published the slices to appropriate nodes. While the [33] developed adaptive partitioning and offloading of multi-end devices to multi-edge nodes based on a matching game approach. Unlike slicing feature map, the authors of [34] sliced the input data into multiple tiles for parallelism. However, both feature-based and input-based partition require frequent synchronous data transmission. Other studies adopt layerwise partition. The authors of [35] extended to the multi-device environment and proposed an iterative algorithm for resource allocation. However, in most cases, model partition degenerates into binary offloading. Thus, the authors of [36] and [37] compressed the intermediate data to reduce the transmission latency with a certain loss of accuracy. In [38], by modeling and solving a batch task scheduling problem, the feasibility of partitioning on a three-exit DNN is also verified under a three-tier framework, i.e., devices, edge, and cloud. In [10], the authors devise a collaborative edge computing system CoopAI to distribute DNN inference over several edge devices with a novel model partition technique to allow the edge devices to prefetch the required data in advance to compute the inference cooperatively in parallel without exchanging data. In [11], the authors propose a technique to divide a DNN in multiple partitions that can be processed locally by end devices or offloaded to one or multiple powerful nodes, such as in fog networks. In [12], the authors propose JointDNN, for collaborative computation between a mobile device and cloud for DNNs in both inference and training phase. In [13], considering online exit prediction and model execution optimization for multi-exit DNN, the authors propose a dynamic path based DNN synergistic inference acceleration framework (DPDS), in which the exit designators are designed to avoid iterative entry for exits. Moreover, the multi-exit DNN is dynamically partitioned according to network environment to achieve fine-grained computing offloading. In [14], the authors design the DNN surgery. The DNN surgery allows partitioned DNN to be processed at both the edge and cloud while limiting the data transmission. Moreover, they design a dynamic adaptive DNN surgery (DADS) scheme to optimally partitions the DNN under different network conditions. In [15], the authors investigate the optimization problem of DNN partitioning in a realistic multiuser resource-constrained condition that rarely considered in previous works. And they propose iterative alternatingoptimization (IAO) algorithm to achieve the optimal solution in polynomial time.

However, even the above works have improved the performance of inference at edge successfully, there is one main shortage that they cannot adaptive the user mobility, i.e., when the user is mobile, the performance of synergistic inference approach between device and edge is deteriorate. Moreover, the above works do not consider the energy consumption of the mobile device, which is much more important than the latency to some extent.

### 2.2 Model segmentation and resource allocation with user mobility

Few works have been proposed to improve the performance of model segmentation and resource allocation under user mobility. For adaptiving the user mobility, the authors in [42] propose a novel collaborative learning scheme for mobile vehicles that can utilize the opportunistic V2R communication to exploit the common priors of vehicular data without interaction with a centralized coordinator. In this approach, vehicles perform local training during the driving journey, and simply upload its local model to RSU encountered on the way. RSU updates the model accordingly and sents back to the vehicle via the V2R communication. Considering that maintaining a satisfactory quality of service when users move across edge servers is challenging, the authors in [18] and [19] propose some novel solutions to improve the quality of inference services for real-time video analytics applications. These approaches includ two schemes: maximizing the use of mobile devices to improve inference quality during the handover and providing offloading decisions to minimize the end-to-end inference delay. The authors in [20] explore on accelerating DNN inference with reliability guarantee in VEC by considering the synergetic impacts of vehicle mobility and V2V/V2I communications. This work shows the necessity of striking a balance between DNN inference acceleration and reliability in VEC and gives insights into the design rationale by analyzing the features of overlapped DNN partitioning and mobility-aware task offloading.

However, the approach in [42] is proposed for mode training, even the approaches in [18], [19] and [20] are proposed for model inference, firstly, the energy consumption and resource limitation are not condiered in these works, secondly, the issues in these works are different from that in this paper, and the technical route in these works cannot sloved the issues proposed in this paper.

# 3  Network Model and Problem Statement

In this section, first, the inference delay model, the energy consumption model, and the resource renting cost model that used in this paper are introduced. The network model is presented in Fig.1. Then, based on the proposed models, the problems that will be solved in this paper are described in detail.

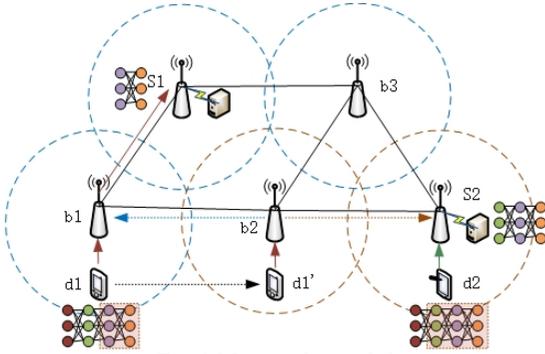

Fig.1 Network model

As shown in Fig. 1, there are $Z$ edge servers and $N$ APs, where $Z < N$. These $Z$ edge servers are deployed on $Z$ APs based on the optimal edge server deployment algorithms, such as [24]. For each AP, it selects one edge server to offload its tasks according to the optimal server selection algorithms [26]. For instantce, in Fig.1, the APs $b_1$ and $b_3$ are served by edge server $S_1$, then the coverage of $S_1$ is the sum of mobile users that covered by $b_1$ and $b_3$. Since one edge server can serve more than one APs, each mobile user can offload tasks to edge server directly or be relayed by other APs. For instance, $d_2$ can offload tasks to $S_2$ directly and $d_1$ needs the help of $b_1$, i.e., $d_1 \rightarrow b_1 \rightarrow S_1$. When the mobile user moves into the coverage of a new edge server, the routing for offloading tasks to edge server will be changed and the task offloading delay is modified, too. Moreover, as depicted in Fig.1, inspired by "data-mule" [39-41], the concept of "model-mule" is proposed, which means that the mobile devices store the whole inference model and move around. When the mobile user moves into the coverage of a new edge server and needs to split the inference model for offloading to edge server, the offloaded part of the inference model is copied and transmitted to edge server. The mobile user still holds the whole model.

## 3.1 Inference delay model

The inference delay includes four parts: 1) the delay that caused by model inference on mobile device; 2) the delay that caused by model inference on edge server; 3) the delay that caused by intermediate data transmission between device and edge server; 4) the delay that caused by the optimal model segmentation and resource allocation strategy calculation.

### A. Inference delay on device

Let the model segmentation decision is $s_i$, which means that the first to $s_i$-th layers are calculated on mobile device $i$, and the $(s_i + 1)$-th to $M$-th layers are offloaded to the edge server for deep inference. $c_i$ is defined as the floating-point operation capability of device $i$. Then, the inference latency on mobile devices after model segmentation can be calculated as:

$$T_i^{device} = \sum_{j=1}^{s_i} \frac{f_{1j}}{c_i} \tag{1}$$

where $f_{1j}$ is the computation task of each layer in the main branch, containing convolutional layer $f_{conv}$, pooling layer $f_{pool}$, and ReLU layer $f_{relu}$ [9]. Thus, $f_{1j}$ can be computed as:

$$f_{1j} = m_{j1}f_{conv} + m_{j2}f_{pool} + m_{j3}f_{relu} \tag{2}$$

where $m_{j1}$, $m_{j2}$, and $m_{j3}$ denote the number of convolutional layer, pooling layer, and ReLU layer, respectively, and $m_{j1} + m_{j2} + m_{j3} = s_i$.

### B. Inference delay on edge server

The execution time is not proportional to the amount of allocated computational resources for the inference tasks, such as DNN, under the scenario that the edge server is multicore CPU. As demonstrated in [15], up to 44% error in execution time between theory and experiment. Thus, in this paper, let $r_i$ represent the number of minimum computational resource unit that allocated to user $i$; $c_{min}$ implies the capability of minimum computational resource unit. Since in the multicore CPU scenario, the execution time is not linear with respect to the amount of allocated computational resource, a compensation function $\lambda(r_i)$ is introduced to fit the exection time in the multicore CPU scenario. For the single core scenario, the $\lambda(r_i)$ is degenerated to $r_i$, and for the multicore scenario, $\lambda(r_i) > r_i$. The $\lambda(r_i)$ can be estimated based on the approach that proposed in [15]. Therefore, in this paper, only assume that $\lambda(r_i)$ increases with $r_i$, but not linear. To model the nonlinearity in the execution time, the execution time on edge can be expressed as:

$$T_i^{server} = \sum_{j=s_i+1}^{M} \frac{f_{ej}}{\lambda(r_i)c_{min}} \tag{3}$$

where $f_{ej}$ is the computation task of each layer in the main branch, containing convolutional layer $f_{conv}$, pooling layer $f_{pool}$, and ReLU layer $f_{relu}$. Thus, $f_{ej}$ can be calculated as:

$$f_{ej} = m_{j4}f_{conv} + m_{j5}f_{pool} + m_{j6}f_{relu} \tag{4}$$

where $m_{j4}$, $m_{j5}$, and $m_{j6}$ denote the number of convolutional layer, pooling layer, and ReLU layer, respectively,

and $m_{j4} + m_{j5} + m_{j6} = M - s_i$.

*C. Network transmission delay*

There are two different kinds of network transmission delay: 1) the intermediate output transmission delay and 2) the final result transmission delay. The delay includes two parts: 1) data transmission delay from end devices to AP and 2) data transmission delay between APs. Firstly, for the data transmission delay from end devices to AP, when the model is splited at $s_i$-th layer, the intermediate data generated by the $s_i$-th layer will be transmitted to edge server to complete the inference. Let $w_{s_i}$ represent the data size at the $s_i$-th layer, $B_i$ is the allocated bandwidth for user $i$, then the intermediate output transmission delay can be expressed as $T_i^{tran-i} = \frac{w_{s_i}}{B_i}$. Let $m_i$ denotes the data size of the final inference result at edge srever, then the final result transmission delay can be expressed by $T_i^{tran-f} = \frac{m_i}{B_i}$.

For the transmission delay between APs, since not all the APs can associate with an edge server, $H_i$ denotes the number of hops from the AP that the mobiles user belongs to to the edge server. Then the delay is $T_i^{relay} = H_i \cdot \left( \frac{w_{s_i}}{B} + \frac{m_i}{B} \right)$, where $B$ means the available bandwidth between APs. Since the APs are connected by optical fiber which has powerful data transmission capability, $B$ represents the bandwidth of available bandwidth in different hops. This is the reason why the effect of bandwidth between APs on reglaying delay is ignored. Thus, in this delay, the effection of data size, the number of hops, and the bandwidth between end devices and AP are taken into consideration.

Thus, the network transmission delay can be expressed as:

$$T_i^{trans} = T_i^{tran-i} + T_i^{tran-f} + T_i^{relay}$$
$$= \frac{w_{s_i} + m_i}{B_i} + H_i \cdot \left( \frac{w_{s_i}}{B} + \frac{m_i}{B} \right) \quad (5)$$

*D. Optimal strategy calculation delay*

In the scenario which considering the user mobility, when the user moves into the coverage of a new edge server, it may need to reculate the optimal model segmentation and resource allocation strategy.

The optimal strategy calculation delay is the time used for calculating the optimal model segmentation and resource allocation strategy, denoted by $T_{Ag}$. However only using $T_{Ag}$ to represent the calculation delay is not fair, because it cannot reflect the real effect on the inference.

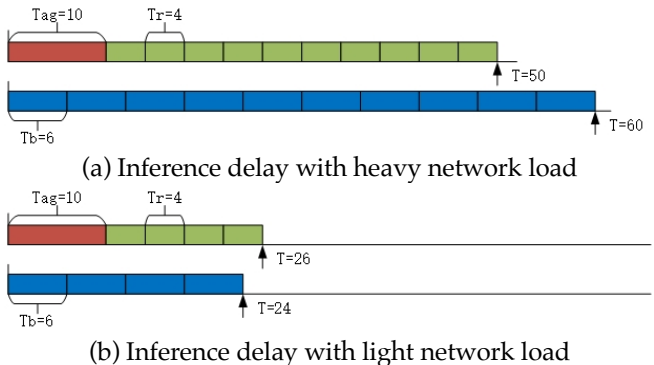

(a) Inference delay with heavy network load

(b) Inference delay with light network load
Fig. 2 Inference delay under different network load

For instance, as demonstrated in Fig.2, the strategy calculation delay of user $i$ is $T_{Ag}^i$ and $T_{Ag}^i = 10$. Moreover, we assume that if recalculating the model segmentation and resource allocation strategy in the new edg server the task calculation time of each round is $T_r^i = 4$, and if transmitting the calculation taks back to the orginal edge server the task calculation time of each round is $T_b^i = 6$. Let $k_i^{high} = 10$ represents the number of task calculation round under heavy workload and $k_i^{low} = 4$ implies the number of task calculation round under light workload. Then, from Fig.2(a), we can find that under the high frequency task calculation, reculating the model segmentation and resource allocation strategy is better than transmitting the task back to the original edge server. Meanwhile, the conclusion is opposite under low frequency task calculation, which is shown in Fig.2(b). Therefore, let $T_{Ag}^i + k_i \cdot T_r^i = T_b^i$, we have:

$$T_{Ag}^i = k_i \cdot \left( T_b^i - T_r^i \right) \quad (6)$$

In (6), the $T_{Ag}^i$, $T_b^i$, and $T_r^i$ are all known once the optimal model segmentation and resource allocation strategy is decided. Therefore, if $k_i$ is large, then the upper limit of $T_{Ag}^i$ is large, too, vice versa. Moreover, the (6) equals to $T_r^i = \frac{T_{Ag}^i}{n} + T_b^i$, this means that for the user who has large number of tasks need to be calculatd on edge server, the effect of convergence time of optimal strategy on task calculation delay is small.

Therefore, in this paper, we define a new parameter, named cost-benefit ratio (CBR), to evaluvate the effect of strategy calculation delay on each round task calculation. The CBR is calculated as:

$$CBR_i = \frac{T_{Ag}^i}{k_i} \quad (7)$$

where $T_{Ag}^i$ is the algorithm convergence time of user $i$; $k_i$ is the number of task calculation round when user $i$ in the coverage of this edge server, i.e., the number of interactions between user $i$ and edge server.

Intuitively, the overall execution latency of the task in mobile user $i$ can be expressed as:

$$T_i = T_i^{device} + T_i^{server} + T_i^{trans} + CBR_i \quad (8)$$

## 3.2 Energy consumption model

Since the energy supply in edge server and AP is ceaseless and enough, we only consider the energy consumption in mobile devices. Additionally, as mentioned before, the energy of mobile devices is all supplied by battery which is always limited on energy store, thus, extending the lifetime of mobile devices is more important than improving the performance of inference. Based on this observation, we need to reduce the energy consumption of mobile devices as much as possible while maintain the inference performance at the same time. The energy consumption that considered in this paper includes: 1) the energy that comsumed for model inference at mobile device and 2) the energy that consumed for intermediate data transmission.

Let $\xi_i$ represent the effective switched capacitance of CPU, which is determined by the chip structure of mobile device $i$, then the energy consumption of mobile devices that used for model inference can be calculated as:

$$E_i^t = \sum_{j=1}^{s_i} \xi_i c_i^2 \varphi_i f_{l_j} \quad (9)$$

where $\varphi_i$ is the required CPU cycles to compute 1-bit data.

Let $p_i$ denotes the transmission power of mobile device $i$, the energy consumption of intermediate data transmission can be computed as:

$$E_i^t = p_i \cdot \frac{w_{s_i} + m_i}{\tau_i} \tag{10}$$

where $\tau_i$ is the data transmission rate and can be obtained by:

$$\tau_i = B_i \cdot \log_2\left(1 + \frac{p_i \alpha_i^\kappa g_i^\kappa}{B_i N_0}\right) \tag{11}$$

where $\alpha_i^\kappa$ implies the large-scale channel fading power component and is negatively related to task transmission distance between mobile device and AP; $g_i^\kappa$ is the small-scale channel fading component.

Thus, the energy consumption for task execution in mobile device and intermediate data transmission between mobile device and edge server can be derived by:

$$\begin{aligned} E_i &= E_i^l + E_i^t \\ &= \sum_{j=1}^{s_i} \xi_i c_i^l \varphi_i f_{l_j} + p_i \cdot \frac{w_{s_i} + m_i}{B_i \cdot \log_2\left(1 + \frac{p_i \alpha_i^\kappa g_i^\kappa}{B_i N_0}\right)} \end{aligned} \tag{12}$$

## 3.3 Resource renting cost model

Except for the energy consumption, the computing resource at edge server and the communication resource at APs are not free for mobile devices. Therefore, we need to investigate the cost model for minimizing the resource renting cost. The resource renting cost comes from two parts: 1) the cost for renting computing resource at edge server and 2) that at APs.

Let $\rho_{min}^j$ represent the renting cost of minimum computational resource unit in edge server $j$, and $r_i^j$ implies the number of minimum computational resource unit in server $j$ that allocated to mobile user $i$, thus, the resource renting cost for model inference at edge can be calculated as:

$$C_{i \to j}^{edge} = r_i^j \rho_{min}^j \tag{13}$$

Additionally, for different edge server, the renting cost is different. The $r_i^j$ relates to the size of model that offloading to edge server. The larger size of model that offloaded to edge server, the more computing resource is needed.

Let $g(*)$ represent the renting cost function of bandwidth, if the allocated bandwidth to mobile user $i$ is $B_i$, then the renting cost can be expressed by:

$$C_i^B = g(B_i) \tag{14}$$

In (14), $g(*)$ is monotone increasing with $B_i$, but not linear. Thus, based on (13) and (14), the cost of resource renting can be calculated as:

$$\begin{aligned} C_i &= C_{i \to j}^{edge} + C_i^B \\ &= r_i^j \rho_{min}^j + g(B_i) \end{aligned} \tag{15}$$

However, the above resource renting cost only considers the size of model and intermediate data, which does not consider the number of task calculation round. Thus, it is not fair. For instance, for mobile device $A$ and $B$, the resource renting cost is $C_A$ and $C_B$, respectively. We assume that $C_A = C_B$ and $k_A > k_B$, where $k$ is the number of task calculation round in mobile device. Then, for each round of task calculation, the resource renting cost in device $A$ is smaller than that in device $B$, even $C_A = C_B$. This means that when the resource renting cost is given, the more tasks are calculated in edge server, the samller resource renting cost for each round task calculation achieves. This is easy to understand since under the same resource renting cost, the more tasks are calculated, the more benefits we can get.

Therefore, according to the definition of CBR in Section 3.1, we define the CBR for resource renting cost, denoted

as $CBR_C^i = \frac{c_i}{k_i}$. Based on (15), $CBR_C^i$ can be expressed as:

$$CBR_C^i = \frac{r_i^j \rho_{min}^j + g(B_i)}{k_i} \tag{16}$$

## 3.4 Problem statement

From the above subsections, we get the inference delay $T_i(s_i, B_i, r_i)$, the energy consumption $E_i(s_i, B_i, r_i)$, and the CBR of resource renting cost $CBR_C^i(s_i, B_i, r_i)$ of mobile user $i$. Our purpose is to achieve minimum inference daly, minimum energy consumption, and minimum resource renting cost at the same time, with the variables are model segmentation strategy $S = \{0,1,2,\ldots,s_i,\ldots,K\}$, communication resource allocation strategy $B = [B_{min} \ B_{max}]$, and computing resource allocation strategy $R = [r_{min} \ r_{max}]$. During these three variables, $S$ is discrete, while $B$ and $R$ are continuous. Therefore, the problem (P0) to be solved in this paper can be expressed as:

$$\min\{\sum_{i=1}^X T_i(s_i, B_i, r_i), \sum_{i=1}^X E_i(s_i, B_i, r_i), \sum_{i=1}^X CBR_C^i(s_i, B_i, r_i)\}$$
$$s.t. \quad 0 \le s_i \le K, \forall i \in [1\ X]$$
$$B_{min} \le B_i \le B_{max}, \forall i \in [1\ X]$$
$$r_{min} \le r_i \le r_{max}, \forall i \in [1\ X]$$

In P0, $X$ represents the number of users. This problem is difficult to be dealt with, since these three optimal objectives are opsite. For instance, for minimizing the resource renting cost, the best approach is calculating the whole inferenece model on device; however, considering the computing capatility of mobile device, this will cause serious inference delay and energy consumption of mobile device. If offload the whole inference model to edge server, the inference delay and the energy consumption of mobile devices will be declined; however, the resource renting cost will be very high. This indicates that finding the minimum energy consumption, minimum inference delay, and minimum resource renting cost at the same time is imposible. Thus, in this paper, we need to find an approach to achieve optimal tradeoff between these three objectives.

# 4 OPTIMAL MODLE SEGMENTATION AND RESOURCE ALLOCATION STRATEGY WITHOUT MOBILITY

In this section, we will investigate the optimal solution for P0 under the scenario in which the end devices are motionless or can move within the coverage of AP. Since P0 is difficult to be solved, we propose an approximate algorithm for P0 in this section. Moreover, in this section, we will investigate the properties of the proposed algorithm.

## 4.1 Loop iteration GD algorithm

Since the optimization objectives shown in P0 are opsite, for constructing the utility for each mobile user that contains both these three objectives, we introduce a weight-based approach to construct the utility function, in which both these three objectives are included. The weight represents the importance of this optimal objective to mobile users. Thus, the utility function of each mobile user can be expressed as:

$$U_i = \omega_T T_i + \omega_C CBR_C^i + \omega_E E_i \tag{17}$$

where $\omega_T$, $\omega_C$, and $\omega_E$ are the weights of inference delay, resource renting cost, and energy consumption, respectively, and $\omega_T + \omega_C + \omega_E = 1$.

Additionally, the $\omega_T$, $\omega_C$, and $\omega_E$ are three hyper-parameters, which can be decided by mobile users according to their multiple and dynamic QoS requirements. For instance, if the inference delay is more important to mobile user than energy consumption and cost, then the mobile user can set $\omega_T > \omega_C$ and $\omega_T > \omega_E$. This approach is flexiable and practicable due to the following reasons. First, in practice, the QoS requirements of the same mobile user may change according to the dynamic environment. For instance, when the mobile devices have enough energy, the inference delay may be the primary factor to be considered; otherwise, when the application requires low inference latency, nomatter whether the energy consumption is enough or the resource renting cost is high or not, the weight of inference delay should be large. Second, for various mobile users and applications, their requirements on QoS may be different. For instance, the energy supply in user $A$ is more than that in user $B$, then the weigh of $\omega_E$ in user $A$ may be smaller than that in user $B$; if one application has strict restriction on inference delay, then the weight of $\omega_T$ in this user could be larger than the other users. Therefore, the weight-based approach can be adjusted according to the dynamic and multiple QoS requirements flexibly.

Thus, according to (17), the P0 can be expressed as:
$$min \sum_{i=1}^{X} U_i$$
Let $U = \sum_{i=1}^{X} U_i$, based on (8), (12) and (16), $U$ can be described as:

$$U = \sum_{i=1}^{X} \omega_T^i T_i(s_i, B_i, r_i) + \sum_{i=1}^{X} \omega_E^i E_i(s_i, B_i, r_i)$$
$$+ \sum_{i=1}^{X} \omega_C^i CBR_i^i(s_i, B_i, r_i)$$
$$= \sum_{i=1}^{X} \omega_T^i \left( \sum_{j=1}^{s_i} \frac{f_{l_j}}{c_i} + \sum_{j=s_i+1}^{M} \frac{f_{e_j}}{\lambda(r_i)c_{min}} + H_i \cdot \frac{w_{s_i}+m_i}{B} + \frac{w_{s_i}+m_i}{B_i} + \frac{T_{dg}^i}{k_i} \right) + \sum_{i=1}^{X} \omega_E^i \left( \sum_{j=1}^{s_i} \xi_i c_i^2 \varphi_i f_{l_j} + p_i \frac{w_{s_i}}{B_i \log_2\left(1 + \frac{p_i \alpha_i^K g_i^K}{B_i N_0}\right)} \right) +$$
$$\sum_{i=1}^{X} \omega_C^i \cdot \frac{r_i^j \rho_{min}^j + g(B_i)}{k_i} \quad (18)$$

In (18), the size of tasks that calculated on mobile device and edge server relates to $s_i$, i.e., $\sum_{j=1}^{s_i} f_{l_j}$ and $\sum_{j=s_i+1}^{M} f_{e_j}$, and $w_{s_i}$ relates to the intermidate data that transmitted between end device and edge server, which cannot be relaxed as continuous variables. Therefore, we define two variables as follows. For user $i$, let $f_i^{i-j} = \sum_{j=1}^{s_i} f_{l_j}$ and $s_i \in \{1, 2, \dots, M\}$, then $f_i^{i-1} = f_{l_1}$, $f_i^{i-2} = f_{l_1} + f_{l_2}$, and so on. Then we can change variable $s_i$ to $f_l^i$, and $f_l^i \in \{f_l^{i-1}, f_l^{i-2}, \dots, f_l^{i-M}\}$, where $f_{l_j}$ is calculated base on (2). Let $Z_i = \sum_{j=1}^{M} f_{l_j}$ is the size of all layers, and $f_e^i = Z_i - f_l^i$. Moreover, $f_l^i$, $f_e^i$, and $w_{s_i}$ are calculated by mobile users in advance and stored in mobile devices with inference model.

Then, we introduce $f_l^i$, $f_e^i$, and $w_{s_i}$ into (18), and get a series of utility functions $U = \{U_1, U_2, \dots, U_{s_i}, \dots, U_M\}$, where $U_{s_i}$ can be expressed as:

$$U_{s_i} = \sum_{i=1}^{X} \omega_T^i \cdot \left( \frac{f_l^i}{c_i} + \frac{f_e^i}{\lambda(r_i)c_{min}} + H_i \cdot \frac{w_{s_i}+m_i}{B} + \frac{w_{s_i}+m_i}{B_i} + \frac{T_{dg}^i}{k_i} \right) +$$

$$\sum_{i=1}^{X} \omega_E^i \left( \xi_i c_i^2 \varphi_i f_l^i + p_i \frac{w_{s_i}}{B_i \log_2\left(1 + \frac{p_i \alpha_i^K g_i^K}{B_i N_0}\right)} \right) + \sum_{i=1}^{X} \omega_C^i \cdot \frac{r_i^j \rho_{min}^j + g(B_i)}{k_i} \quad (19)$$

where $f_l^i$, $f_e^i$, and $w_{s_i}$ are already known in advance for each inference model in mobile device; $B_i \in [B_{min} \ B_{max}]$ and $r_i \in [r_{min} \ r_{max}]$.

In (19), on the one hand, there are $X$ mobile users. For each mobile user, we need to calculate the optimal $B$ and $r$. Moreover, since the parameters $B$ and $r$ are continuous, the variable spaces of $B$ and $r$ are large and infinite-dimensional. Additionally, $B$ and $r$ are all related to $s_i$, so it is difficult to calculate the optimal value of $B$ and $r$ separately. Thus, in this paper, for finding the optimal tradeoff between inference delay, energy consumption, and resource renting cost, we introduce the gradient descent approach into our algorithm. For using the gradient descent to address above issue, firstly, we need to prove that the weight function shown in (18) is differentiable. The definition of differentiable is introduced as follows.

**Definition 1.** For $f(x, y, z)$, if its partial derivative on $x$, $y$, and $z$, i.e., $f'(x, y, z)|_x$, $f'(x, y, z)|_y$, and $f'(x, y, z)|_z$, exist and continue, the $f(x, y, z)$ is differentable.

Based on Definition 1, we have the conclusion as follows.
**Corollary 1.** When the values of $f_l^i$, $f_e^i$, and $w_{s_i}$ are know in advance, the utility function shown in (18) is differentiable.

*Proof.* Let $y_1 = \frac{f_l^i}{c_i} + H_i \cdot \frac{w_{s_i}+m_i}{B} + \frac{T_{dg}^i}{k_i}$ and $y_2 = \xi_i c_i^2 \varphi_i f_l^i$, then, the (18) can be expressed as:

$$U_{s_i} = \sum_{i=1}^{X} \omega_T^i \cdot \left( y_1 + \frac{f_e^i}{\lambda(r_i)c_{min}} + \frac{w_{s_i}+m_i}{B_i} \right) + \sum_{i=1}^{X} \omega_E^i \cdot \left( y_2 + p_i \frac{w_{s_i}}{B_i \log_2\left(1 + \frac{p_i \alpha_i^K g_i^K}{B_i N_0}\right)} \right) + \sum_{i=1}^{X} \omega_C^i \cdot \frac{r_i^j \rho_{min}^j + g(B_i)}{k_i} \quad (20)$$

Since the values of $f_l^i$, $f_e^i$, and $w_{s_i}$ are already know in advance for every inference model in each mobile device, the values of $y_1$ and $y_2$ can be calculated easily and nothing to do with the partial derivative on $B_i$ and $r_i$. Therefore, in (20), the variables are $B_i$ and $r_i$. Thus, the partial derivative of $U_{s_i}$ on $B_i$ and $r_i$ can be expressed as:

$$U_{s_i}'|_{B_i} = \sum_{i=1}^{X} \omega_T^i \cdot \left( -\frac{w_{s_i}+m_i}{(B_i)^2} \right) + \sum_{i=1}^{X} \omega_E^i p_i w_{s_i} \cdot$$
$$\left( -\frac{\log_2\left(1 + \frac{p_i \alpha_i^K g_i^K}{B_i N_0}\right) - \frac{p_i \alpha_i^K g_i^K}{B_i N_0\left(1 + \frac{p_i \alpha_i^K g_i^K}{B_i N_0}\right)\ln 2}}{\left(B_i \log_2\left(1 + \frac{p_i \alpha_i^K g_i^K}{B_i N_0}\right)\right)^2} \right) + \sum_{i=1}^{X} \omega_C^i \cdot g'(B_i) \quad (21)$$

$$U_{s_i}'|_{r_i} = \sum_{i=1}^{X} \omega_T^i \cdot \frac{f_e^i}{c_{min}} \cdot \left( -\frac{\lambda'(r_i)}{(\lambda(r_i))^2} \right) + \sum_{i=1}^{X} \omega_C^i \cdot \rho_{min}^j \quad (22)$$

For $U_{s_i}'|_{B_i}$ shown in (21), the $B_i \in [B_{min} \ B_{max}]$ and $B_{min} > 0$. Then, for $\forall B_i \in [B_{min} \ B_{max}]$, we have $x_1 = -\frac{w_{s_i}+m_i}{(B_i)^2}$ is continuous, $x_2 = B_i \log_2\left(1 + \frac{p_i \alpha_i^K g_i^K}{B_i N_0}\right)$ is continuous, $x_3 = \log_2\left(1 + \frac{p_i \alpha_i^K g_i^K}{B_i N_0}\right)$ is continuous, and $x_4 = g'(B_i)$ is continuous. Thus, based on the operational rule of continuous function, the $U_{s_i}'|_{B_i}$ is continuous with $\forall B_i \in [B_{min} \ B_{max}]$. Similarly, since $\lambda(r_i)$ is continuous, $U_{s_i}'|_{r_i}$ is continuous with $\forall r_i \in [r_{min} \ \ r_{max}]$. Therefore, Corollary 1 is proved. ∎

The Corollary 1 means that when the values of $f_i^t$, $f_e^i$, and $w_{s_i}$ are know, the GD approach can be used in (19) to find the optimal strategies of $B$ and $r$. However, the utility function shown in (19) is only the utility when the model segmentation point is $s_i$, there are $M$ layers in the inference model, which means that the GD algorithm needs to be repeated $M$ times to find the global optimal solutions for $B$ and $r$. However, considering the complexity and convergence time of the GD approach, repeating the GD approach $M$ times will cause serious delay and complexity. Fortunately, for the GD approach, if we can select the initial value carefully, the complexity and convergence time will be reduced greatly. Therefore, in this paper, based on the greedy approach, we propose the Loop iteration GD algorithm, shorted as Li-GD. The details of the Li-GD algorithm are presented in Table 1.

Table 1

| **Algorithm 1:** Loop iteration GD algorithm (Li-GD) |
|---|
| **Input:** |
|     Objective function: $\boldsymbol{U} = \{U_1, U_2, \dots, U_{s_i}, \dots, U_M\}$; |
|     Gradient function: $\boldsymbol{\nabla} = \{\nabla_{B_i} = \frac{\partial u_{s_i}}{\partial B_i}, \nabla_{r_i} = \frac{\partial u_{s_i}}{\partial r_i}\}$; |
|     Algorithm accuracy: $\varepsilon$; |
|     Step size: $\lambda$; |
| **Output:** |
|     The optimal solution $\boldsymbol{O}^* = \{\boldsymbol{B}^*, \boldsymbol{r}^*\}$; |
| 1.   Let $\boldsymbol{B}^{j(0)} \in R$ and $\boldsymbol{r}^{j(0)} \in R$, $\forall i \in [1 \ X]$ and $\forall j \in [1 \ M]$; |
| *# Calculating the optimal strategy for the first layer #* |
| 2.   If $j = 1$; |
| 3.     Let $k \leftarrow 0$, $\boldsymbol{B}^{j(k)} = \{B_1^{j(k)}, B_2^{j(k)}, \dots, B_M^{j(k)}\}$ and $\boldsymbol{r}^{j(k)} = \{r_1^{j(k)}, r_2^{j(k)}, \dots, r_M^{j(k)}\}$; |
| 4.     Calculating $U_{s_i}(\boldsymbol{B}^{j(k)}, \boldsymbol{r}^{j(k)})$; |
| 5.     Calculating the gradient $\boldsymbol{g}_k = g(\boldsymbol{B}^{j(k)}, \boldsymbol{r}^{j(k)})$; |
| 6.     If $\|g_k\| < \varepsilon$, then $\boldsymbol{B}^{j*} \leftarrow \boldsymbol{B}^{j(k)}$ and $\boldsymbol{r}^{j*} \leftarrow \boldsymbol{r}^{j(k)}$; |
| 7.     Otherwise, let $\boldsymbol{p}_k = -g(\boldsymbol{B}^{j(k)}, \boldsymbol{r}^{j(k)})$, and let $\boldsymbol{B}^{j(k+1)} = \boldsymbol{B}^{j(k)} + \lambda \boldsymbol{p}_k$ and $\boldsymbol{r}^{j(k+1)} = \boldsymbol{r}^{j(k)} + \lambda \boldsymbol{p}_k$; |
| 8.       Calculating $U_{s_i}(\boldsymbol{B}^{j(k+1)}, \boldsymbol{r}^{j(k+1)}) = U_{s_i}(\boldsymbol{B}^{j(k)} + \lambda \boldsymbol{p}_k, \boldsymbol{r}^{j(k)} + \lambda \boldsymbol{p}_k)$; |
| 9.     If $\|U_{s_i}(\boldsymbol{B}^{j(k+1)}, \boldsymbol{r}^{j(k+1)}) - U_{s_i}(\boldsymbol{B}^{j(k)}, \boldsymbol{r}^{j(k)})\| < \varepsilon$ or $\max\{\|\boldsymbol{B}^{j(k+1)} - \boldsymbol{B}^{j(k)}\|, \|\boldsymbol{r}^{j(k+1)} - \boldsymbol{r}^{j(k)}\|\} < \varepsilon$; |
| 10.       then $\boldsymbol{B}^{j*} \leftarrow \boldsymbol{B}^{j(k+1)}$ and $\boldsymbol{r}^{j*} \leftarrow \boldsymbol{r}^{j(k+1)}$; |
| 11.     otherwise, $k = k + 1$; |
| 12.   end if |
| *# Calculating the optimal strategy of the rest layers #* |
| 13.   When $1 < j \le M$; |
| *# Loop iteration #* |
| 14.     Let $\boldsymbol{B}^{j+1(0)} = \boldsymbol{B}^{j*}$ and $\boldsymbol{r}^{j+1(0)} = \boldsymbol{r}^{j*}$, $\forall i \in [1 \ X]$ and $\forall j \in [1 \ M]$; |
| 15.     repeating step 3 to Step 11; |
| 16.     $j = j + 1$; |
| *# Finding the optimal strategy #* |
| 17.   Calculating $\boldsymbol{U} = \{U_1(\boldsymbol{B}^{1*}, \boldsymbol{r}^{1*}), U_2(\boldsymbol{B}^{2*}, \boldsymbol{r}^{2*}), \dots, U_M(\boldsymbol{B}^{M*}, \boldsymbol{r}^{M*})\}$; |
| 18.   $(\boldsymbol{s}, \boldsymbol{B}, \boldsymbol{r}) \leftarrow \arg \min_{\boldsymbol{s}^*, \boldsymbol{B}^*, \boldsymbol{r}^*} \boldsymbol{U}$. |

The Li-GD algorithm presented in Table 1 is composed of three parts.

1) (Line2-Line12): Calculating the optimal strategy when the model segmentation point is in the first layer. For the Li-GD algorithm, since the model segmentation strategy is discrete, we need to calculate the optimal resource allocation strategy layer by layer. For the first layer, its starting values are $\boldsymbol{B}^{1(0)} = \{B_1^{1(0)}, B_2^{1(0)}, \dots, B_X^{1(0)}\}$ and $\boldsymbol{r}^{1(0)} = \{r_1^{1(0)}, r_2^{1(0)}, \dots, r_X^{1(0)}\}$, where $\boldsymbol{B}^{j(0)} \in R$ and $\boldsymbol{r}^{j(0)} \in R$, and they are selected without any information of the final optimal values. Then, the GD algorithm is excuted with step size of $\lambda$ and gradient of $-\boldsymbol{g}_k$. After $k$ rounds iteration, when the threshold of accuracy is reached, the optimal solutions of resource allocation strategy for the first layer are $\boldsymbol{B}^{1*} \leftarrow \boldsymbol{B}^{1(k)}$ and $\boldsymbol{r}^{1*} \leftarrow \boldsymbol{r}^{1(k)}$.

2) (Line13-Line16): Calculating the optimal resource strategy for the rest layers. When the optimal resource allocation strategy of the first layer is calculated, then from the second layer, the starting values of this layer is the optimal values of the last layer. For instance, $\boldsymbol{B}^{2(0)} = \boldsymbol{B}^{1*}$ and $\boldsymbol{r}^{2(0)} = \boldsymbol{r}^{1*}$, $\boldsymbol{B}^{3(0)} = \boldsymbol{B}^{2*}$ and $\boldsymbol{r}^{3(0)} = \boldsymbol{r}^{2*}$, etc. The GD process is the same as that when calculating the optimal resource allocation strategy for first layer. In this stage, the optimal resource allocation strategies for $2th$ layer to the $Mth$ layer are calculated, which are $\boldsymbol{B}^{(2-M)*} = \{\boldsymbol{B}^{2*}, \dots, \boldsymbol{B}^{M*}\}$ and $\boldsymbol{r}^{(2-M)*} = \{\boldsymbol{r}^{2*}, \dots, \boldsymbol{r}^{M*}\}$.

3) (Line17-Line18): Finding the final optimal model segmentation strategy and resource allocation strategy. When the optimal resource allocation strategies for all the layers are calculated, which are $\boldsymbol{B}^* = \{\boldsymbol{B}^{1*}, \boldsymbol{B}^{2*}, \dots, \boldsymbol{B}^{M*}\}$ and $\boldsymbol{r}^* = \{\boldsymbol{r}^{1*}, \boldsymbol{r}^{2*}, \dots, \boldsymbol{r}^{M*}\}$, respectively, then substituting the $\boldsymbol{B}^*$ and $\boldsymbol{r}^*$ into (18) and getting $M$ utility values $\boldsymbol{U}^* = \{\boldsymbol{U}^{1*}, \boldsymbol{U}^{2*}, \dots, \boldsymbol{U}^{M*}\}$. Finally, finding the minimum value from $\boldsymbol{U}^*$, and the model segmentation strategy and resource allocation strategy that associated with this utility value is seleced as the final optimal strategy.

The theory foundations of Li-GD approach are: 1) for the GD algorithm, carefully selecting the start value can decrease the complexity and speed up the convergence greatly [43]; 2) for the inference model, in most cases, the optimal resource allocation strategy between adjacent layers is much more similarly than the other laysers. Since the size of layers and the intermediate transmission data between adjacent lyaers are similar [9]. For proving the effectiveness of the Li-GD algorithm that proposed in this section, we give the conclusions as follows.

## 4.2 The properties of Li-GD algorithm

In this section, we will investigate the properties of Li-GD algorithm, including convergence and complexity. The details are shown below.

**Corollary 2.** The Li-GD algorithm is convergent, and the convergence time is $K = \frac{\|x^0 - x^*\|_2^2}{2\eta\epsilon}$, where $\eta$ is the step size and $\eta \le \frac{1}{L}$, $\epsilon$ is the threshold of accuracy.

*Proof.* Based on the conclusions in [43], if the difference-able function $f(x)$ satisfies: 1) L-Lipschitz smooth and 2) convex, the $f(x)$ is convergent.

Based on (20), the parts of $U_{s_i}$ relate to $B_i$ are:

$$U_{s_i}(B_i) = \sum_{i=1}^{X} \omega_T^i \cdot \frac{w_{s_i} + m_i}{B_i}$$
$$+ \sum_{i=1}^{X} \omega_E^i \cdot p_i \frac{w_{s_i}}{B_i \log_2\left(1 + \frac{p_i \alpha_i^k g_i^k}{B_i N_0}\right)} + \sum_{i=1}^{X} \omega_C^i \cdot \frac{g(B_i)}{k_i} \quad (23)$$

For $U_{s_i}(B_i)$, based on the conclusions in [43], the first term satisifies the above three conditions. Moreover, we have assumed that the $g(B_i)$ also satisifies the above three conditions. The second term is complex. In the following, we will discuss the second term in detail.

The second term in (23) can be simplified as:

$$f(x) = \frac{1}{x \log_2\left(1 + \frac{1}{x}\right)} \quad (24)$$

And the first-order derivative of (24) can be expressed as:

$$f'(x) = \frac{1}{x^2 \log_2\left(1 + \frac{1}{x}\right)} \left(\frac{1}{(1+x)\ln 2 \log_2\left(1 + \frac{1}{x}\right)} - 1\right) \quad (25)$$

*L-Lipschitz smooth:*

Let $y = kx$ and $k > 1$, then we have $|x - y| = x|k - 1|$ and:

$$\left|f'^{(x)} - f'(y)\right| = \left| \frac{1}{x^2 \log_2\left(1 + \frac{1}{x}\right)} \left(\frac{1}{(1+x)\ln 2 \log_2\left(1 + \frac{1}{x}\right)} - 1\right) - \frac{1}{kx^2 \log_2\left(1 + \frac{1}{kx}\right)} \left(\frac{1}{(1+kx)\ln 2 \log_2\left(1 + \frac{1}{kx}\right)} - 1\right) \right| \quad (26)$$

Since $\frac{1}{\log_2\left(1 + \frac{1}{x}\right)}$ increases with the increasing of $x$, then we have:

$$|f'(x) - f'(y)| \leq \left| \frac{1}{kx^2 \log_2\left(1 + \frac{1}{x}\right)} \left(\frac{1}{(1+kx)\ln 2 \log_2\left(1 + \frac{1}{x}\right)} - 1\right) - \frac{1}{x^2 \log_2\left(1 + \frac{1}{x}\right)} \left(\frac{1}{(1+x)\ln 2 \log_2\left(1 + \frac{1}{x}\right)} - 1\right) \right|$$
$$= \frac{1}{x^2 \log_2\left(1 + \frac{1}{x}\right)} \left| \frac{1}{k(1+kx)\ln 2 \log_2\left(1 + \frac{1}{x}\right)} - \frac{1}{(1+x) \log_2\left(1 + \frac{1}{x}\right)} \right|$$
$$= \frac{1}{\left[x \log_2\left(1 + \frac{1}{x}\right)\right]^2 \ln 2} \left| \frac{1}{k(1+kx)} - \frac{1}{(1+x)} \right|$$
$$\leq \frac{1}{\left[x \log_2\left(1 + \frac{1}{x}\right)\right]^2 \ln 2} \left| \frac{1}{k(1+x)} - \frac{1}{(1+x)} \right|$$
$$= \frac{1}{(1+x)\left[x \log_2\left(1 + \frac{1}{x}\right)\right]^2 \ln 2} \left| 1 - \frac{1}{k} \right| < Lx|k - 1| = L|x - y| \quad (27)$$

Moreover, since $L > \frac{\frac{1}{(1+x)\left[x \log_2\left(1 + \frac{1}{x}\right)\right]^2 \ln 2} \left| 1 - \frac{1}{k} \right|}{x|k-1|}$, we have:

$$L > \frac{1}{k \ln 2} \cdot \frac{1}{x^3(1+x)\left[\log_2\left(1 + \frac{1}{x}\right)\right]^2} \quad (28)$$

Since $\frac{1}{x^3(1+x)\left[\log_2\left(1 + \frac{1}{x}\right)\right]^2} < 1$, then we can conclude that the lower bound of $L$ is samller than $\frac{1}{k \ln 2}$.

*Convex:*

For the differenceable function $f(x)$, if $f''(x) > 0$, then it is convex [43]. The $f'(x)$ is presented in (25). Thus, let $A = \frac{1}{x^2 \log_2\left(1 + \frac{1}{x}\right)}$, $B = \frac{1}{(1+x)\ln 2 \log_2\left(1 + \frac{1}{x}\right)} - 1$, then we have:

$$A' = \frac{1}{x^3 \log_2\left(1 + \frac{1}{x}\right)} \left[\frac{1}{(1+x)\ln 2 \log_2\left(1 + \frac{1}{x}\right)} - 2\right] \quad (29)$$

$$B' = \frac{1}{(1+x)^3 \ln 2 \log_2\left(1 + \frac{1}{x}\right)} \left[\frac{1}{x \ln 2 \log_2\left(1 + \frac{1}{x}\right)} - 1\right] \quad (30)$$

Therefore,

$$f''(x) = \frac{1}{x^3 \log_2\left(1 + \frac{1}{x}\right)} \left[\frac{1}{(1+x)\ln 2 \log_2\left(1 + \frac{1}{x}\right)} - 2\right] \cdot \left[\frac{1}{(1+x)\ln 2 \log_2\left(1 + \frac{1}{x}\right)} - 1\right] + \frac{1}{(1+x)^3 \ln 2 \log_2\left(1 + \frac{1}{x}\right)} \left[\frac{1}{x \ln 2 \log_2\left(1 + \frac{1}{x}\right)} - 1\right] \cdot$$

$$\frac{1}{x^2 \log_2\left(1 + \frac{1}{x}\right)} \quad (31)$$

Since $\left[\frac{1}{(1+x)\ln 2 \log_2\left(1 + \frac{1}{x}\right)} - 2\right] < \left[\frac{1}{(1+x)\ln 2 \log_2\left(1 + \frac{1}{x}\right)} - 1\right]$, then we have

$$f''(x) < \frac{1}{x^3 \log_2\left(1 + \frac{1}{x}\right)} \left[\frac{1}{(1+x)\ln 2 \log_2\left(1 + \frac{1}{x}\right)} - 2\right]^2 + \frac{1}{(1+x)^3 \ln 2 \log_2\left(1 + \frac{1}{x}\right)} \left[\frac{1}{x \ln 2 \log_2\left(1 + \frac{1}{x}\right)} - 1\right] \cdot \frac{1}{x^2 \log_2\left(1 + \frac{1}{x}\right)} \quad (32)$$

Thus, if $g(x) = \frac{1}{x \ln 2 \log_2\left(1 + \frac{1}{x}\right)} - 1 > 0$, the function $f(x)$ is convex. Moreover, $g(x) > 0$ equals to:

$$\frac{1}{x \ln 2 \log_2\left(1 + \frac{1}{x}\right)} > 1 \quad (33)$$

$$\log_2\left(1 + \frac{1}{x}\right) < \frac{1}{x \ln 2} \quad (34)$$

Based on the Bottoming formula, we have:

$$\log_2\left(1 + \frac{1}{x}\right) < \log_2 e^{\frac{1}{x}} \quad (35)$$

which equals to:

$$\left(1 + \frac{1}{x}\right) < e^{\frac{1}{x}} \quad (36)$$

Since the Taylor expansion of $e^{\frac{1}{x}}$ is:

$$e^{\frac{1}{x}} = \sum_{n=0}^{\infty} \frac{(1/x)^n}{n!} \quad (37)$$

which is larger than $\left(1 + \frac{1}{x}\right)$. Thus, the $f(x)$ is convex.

Based on (20), the parts of $U_{s_i}$ relate to $r_i$ are:

$$U_{s_i} = \sum_{i=1}^{X} \omega_T^i \cdot \frac{f_c^i}{\lambda(r_i)c_{min}} + \sum_{i=1}^{X} \omega_C^i \cdot \frac{r_i^j \rho_{min}^j}{k_i} \quad (38)$$

For (38), the second term is L-Lipschitz smooth and convex. For the first term in (38), we assume that the $\lambda(r_i)$ is L-Lipschitz smooth and convex, thus, for $U_{s_i}(B_i)$, we can prove that the first term in (38) is also L-Lipschitz smooth and convex. Therefore, the Corollary 2 is proved.∎

**Corollary 3.** The average complex of the Li-GD algorithm is $O(X\overline{K}Mx^3 \ln^2(x))$, where $\overline{K} = \frac{\sum_{i=1}^{M} K^i}{M}$.

*Proof.* Based on the Corollary 2, the convergence time of Li-GD approach is $K = \frac{\|x^0 - x^*\|_2^2}{2\eta \epsilon}$. In Li-GD algorithm, for each round of iteration, we need to calculate the gradient for $B_i$ and $r_i$, and the number of mobile users is $X$. Moreover, the gradient calculation of $B_i$, which is shown in (25), is much more complex than $r_i$. The (25) can be rewritten as:

$$f'(x) = \frac{1}{x^2 \log_2\left(1 + \frac{1}{x}\right)} \left(\frac{1}{(1+x)\ln 2 \log_2\left(1 + \frac{1}{x}\right)} - 1\right)$$
$$= \frac{1}{x^2 \log_2\left(1 + \frac{1}{x}\right)} \cdot \frac{1}{(1+x)\ln 2 \log_2\left(1 + \frac{1}{x}\right)} - \frac{1}{x^2 \log_2\left(1 + \frac{1}{x}\right)} \quad (39)$$

Since the first term is more complex than the second term, we only consider the first term, let:

$$h'(x) = \frac{1}{x^2 \log_2\left(1 + \frac{1}{x}\right)} \cdot \frac{1}{(1+x)\ln 2 \log_2\left(1 + \frac{1}{x}\right)}$$
$$= \frac{1}{x^2(1+x)\left[\log_2\left(1 + \frac{1}{x}\right)\right]^2 \ln 2} = \frac{1}{(x^2+x^3)[\log_2(1+x) - \log_2(x)]^2 \ln 2}$$
$$= O(x^3 \ln^2(x)) \quad (40)$$

Moreover, for each mobile user in each round of GD algorithm, the (40) should be calculated. Additionally, the number of layers in the Li-GD algorithm is $M$. Thus, the worst case complexity of the Li-GD algorithm that proposed in this paper is $O(XK^*Mx^3 \ln^2(x))$, where $K^*$ is the maximum convergence time during $M$ layers and $K^* = \max\{K^1, K^2, ..., K^M\}$. The average complexity of the Li-GD algorithm is $O(X\overline{K}Mx^3\ln^2(x))$, where $\overline{K} = \frac{\sum_{i=1}^{M}K^i}{M}$. ∎

**Corollary 4.** The Li-GD algorithm can accelerate the convergence of GD algorithm while reducing complexity.

*Proof.* From Corollary 2 and Corollary 3, we can find that for the fixed precision $\varepsilon$ and step size $\eta$, the convergence time $K$ is corelated to the starting point $\boldsymbol{B}^{(0)}$ and $\boldsymbol{r}^{(0)}$, and the complexity is associated with the convergence time $K$. Thus, for reducing convergence time and complexity, the starting point should be selected carefully.

In the traditional GD algorithm, for each round of GD, the starting values are $\boldsymbol{B}^{(0)}$ and $\boldsymbol{r}^{(0)}$. The convergence time is $K_1 = \max\left\{\frac{\|\boldsymbol{B}^{(0)}-\boldsymbol{B}^*\|_2^2}{2\eta\varepsilon}, \frac{\|\boldsymbol{r}^{(0)}-\boldsymbol{r}^*\|_2^2}{2\eta\varepsilon}\right\}$. However, as introduced in Section 4.1, for the Li-GD algorithm, the starting values are the optimal solutions of last layer, which are $\boldsymbol{B}^{j+1(0)} = \boldsymbol{B}^{j*}$ and $\boldsymbol{r}^{j+1(0)} = \boldsymbol{r}^{j*}$. Therefore, the convergence time of Li-GD algorithm is $K_2 = \max\left\{\frac{\|\boldsymbol{B}^{j+1(0)}-\boldsymbol{B}^{j*}\|_2^2}{2\eta\varepsilon}, \frac{\|\boldsymbol{r}^{j+1(0)}-\boldsymbol{r}^{j*}\|_2^2}{2\eta\varepsilon}\right\}$. Since $|\boldsymbol{B}^{j+1(0)}-\boldsymbol{B}^{j*}|$ and $|\boldsymbol{r}^{j+1(0)}-\boldsymbol{r}^{j*}|$ are much smaller than $|\boldsymbol{B}^{(0)}-\boldsymbol{B}^*|$ and $|\boldsymbol{r}^{(0)}-\boldsymbol{r}^*|$, the convergence time is accelerated. ∎

Additionally, there are $M$ layers, for the traditional GD algorithm, the total convergence time is $MK_1$. For Li-GD algorithm, the total convergence time is $K_1 + \sum_{j=2}^{M}K_2^j$. Since $K_2^j$ is much smaller than $K_1$, the complexity is declined. ∎

Note that the convergence time can be reduced further by optimizing the step size or using self-adaptive step size. Moreover, lowering the accuracy can also accelerate the convergence. Therefore, by carefully achieving tradeoff between accuracy and convergence time also can improve the performance of Li-GD algorithm. However, these are not investigated in this paper.

# 5 OPTIMAL MODLE SEGMENTATION AND RESOURCE ALLOCATION STRATEGY WITH MOBILITY

In Section 4, we have investigated the optimal model segmentation and resource allocation strategy without user mobility. However, for the scenario in which the users can move, the calculation of the optimal solution becomes more difficult. As demonstrated in Fig.1, when user $d_1$ moves from the coverage of $s_1$ to the coverage of $s_2$, it has three alternative stragegies based on "mode-mule": 1) reculating the optimal model segmentation strategy between user $d_1$ and $s_2$, 2) transmitting the inference task back to $s_1$, or 3) migrating the offloaded model in $s_1$ to $s_2$. However, on the one hand, since the size of model on $s_1$ is large, migrating it to $s_1$ will cause serious delay, and the delay is higher than the intermediate data transmission delay in the strategy that transmitting the inference task back to $s_1$; on the other hand, considering the heterogeneity of edge servers and the changing of offloading path, the optimal model segmentation between $d_1$ and $s_1$ may not be optimal between $d_1$ and $s_2$ any more. Therefore, the inference delay may be higher than the strategy which recalculating the model segmentation and resource allocation. In consequence, we only considering two alternative stratigies in this apper: 1) reculating the optimal model segmentation strategy between user $d_1$ and $s_2$, and 2) transmitting the inference task back to

$s_1$.

Let "0" represent the strategy that reculating the optimal model segmentation and "1" denote the stratetegy that transmitting the inference task back to the original edge server, according to the conclusions in Section 4, we have:

$$\boldsymbol{U} = (\mathbf{1} - \boldsymbol{R})\boldsymbol{U_1} + \boldsymbol{R}\boldsymbol{U_2} \quad (41)$$

where $\boldsymbol{R}$ is the set of decision for $X$ mobile users and $\boldsymbol{R} = \{R_1, R_2, ..., R_X\}$ with $R_i = \{0,1\}$ for $\forall i \in [1\ X]$; $\boldsymbol{U_1}$ is the set of utility for reculating the optimal model segmentation of $X$ mobile users and $\boldsymbol{U_1} = \{U_1^1, U_1^2, ..., U_1^X\}$; $\boldsymbol{U_2}$ is the set of utility for transmitting the inference task back to the original edge server of $X$ mobile users and $\boldsymbol{U_2} = \{U_2^1, U_2^2, ..., U_2^X\}$; $\mathbf{1}$ is the all "1" matrix with the same dimensions as $\boldsymbol{R}$. The $U_1$ can be calculated based on (18). The $U_2^i$ is the utility of user $i$ when transmitting the inference task back to the original server and it can be divided into three parts: 1) the energy consumption, inference delay, and resource renting cost of mobile device, denoted as $U_2^{id}$; 2) the energy consumption, inference delay, and resource renting cost of edge server, denoted as $U_2^{ie}$; 3) the intermediate data and final result transmission delay $U_2^{iv}$. The $U_2^{id}$ and $U_2^{ie}$ keep constant when the mobile user transmits the inference task back to the original edge server. Only the $U_2^{iv}$ varies due to the changing of the task offloading and final result transmission path. Additionally, $U_2^{iv}$ can be calculated as:

$$U_2^{iv} = \frac{w_{s_i}+m_i}{B_i^2} + H_i^2 \cdot \frac{w_{s_i}+m_i}{B} \quad (42)$$

where $H_i^2$ is the number of hops from mobile user to edge server, $B_i^2$ is the allocated bandwidth of mobile users in the new AP. The $H_i^2$ is calculated based on Dijkstra algorithm. Thus, (41) becomes:

$$U_{s_i} = (\mathbf{1} - \boldsymbol{R}) \cdot \left\{ \sum_{i=1}^{X}\omega_T^i\left(\sum_{j=1}^{s_i}\frac{f_{l_j}}{c_i} + \sum_{j=s_i+1}^{M}\frac{f_{e_j}}{\lambda(r_i)c_{min}} + H_i \cdot \right. \right.$$

$$\frac{w_{s_i}+m_i}{B} + \frac{w_{s_i}+m_i}{B_i} + \frac{T_{dg}^i}{k_i}\right) + \sum_{i=1}^{X}\omega_E^i\left(\sum_{j=1}^{s_i}\xi_i c_i^2 \varphi_i f_{l_j} + \right.$$

$$\left. p_i\frac{w_{s_i}}{B_i\log_2\left(1+\frac{p_i\alpha_i^K g_i^K}{B_i N_0}\right)}\right) + \sum_{i=1}^{X}\omega_C^i \cdot \frac{r_i^j\rho_{min}^j+g(B_i)}{k_i}\right\} + \boldsymbol{R} \cdot$$

$$\left\{\sum_{i=1}^{X}\left(\frac{w_{s_i}^i+m_i^i}{B_i^2} + H_i^2 \cdot \frac{w_{s_i}^i+m_i^i}{B}\right) + \sum_{i=1}^{X}\left(U_2^{id}+U_2^{ie}\right)\right\} \quad (43)$$

Since $R = \{0,1\}$, the (43) is not differentiable. Thus, we relax $R = \{0,1\}$ to $R \in [0\ 1]$. The calculation of (43) is also the Li-GD algorithm. The differences between (18) and (43) are: 1) the vairables in (43) are $R_i$, $B_i$, and $r_i$, which are more complex than (18); 2) since $R = \{0,1\}$ is relaxed to $R \in [0\ 1]$, the solution of $R_i$ is approximate. In (43), the first term will be traversed from the 1st layer to the $M$ th layer of the inference model; however, since the model segmentation strategy in the second term does not change, the size of the intermediate data and the final result is fixed. Only the resource allocation strategy and data transmission path in the second term varies with different APs. For calculating the optimal solutions of (43), we propose the mobility aware Li-GD (MLi-GD) algorithm in Table 2. The details of Algorithm 2 are shown as follows.

Table 2

| Algorithm 2: Mobility aware Li-GD algorithm |
| --- |
| **Input:** |
| Objective function: $\boldsymbol{U} = \{U_1, U_2, ..., U_{s_i}, ..., U_M\}$; |

Gradient function: $\nabla = \{\nabla_{B_i} = \frac{\partial U_{s_i}}{\partial B_i}, \nabla_{r_i} = \frac{\partial U_{s_i}}{\partial r_i}, \nabla_{R_i} = \frac{\partial U_{s_i}}{\partial R_i}\}$;

Algorithm accuracy: $\varepsilon$;

Step size: $\lambda$;

**Output:**

The optimal solution $\boldsymbol{O^*} = \{\boldsymbol{B^*}, \boldsymbol{r^*}, \boldsymbol{R^*}\}$;

1. Let $\boldsymbol{B}^{j(0)} \in R$, $\boldsymbol{r}^{j(0)} \in R$, and $\boldsymbol{R}^{j(0)} \in R$, $\forall i \in [1\ X]$ and $\forall j \in [1\ M]$;

*# Calculating the optimal strategy for the first layer #*

2. If $j = 1$;

3. Let $k \leftarrow 0$, $\boldsymbol{B}^{j(k)} = \left\{B_1^{j(k)}, B_2^{j(k)}, \ldots, B_M^{j(k)}, B_{si}^{j(k)}\right\}$, $\boldsymbol{R}^{j(k)} = \left\{R_1^{j(k)}, R_2^{j(k)}, \ldots, R_M^{j(k)}\right\}$, and $\boldsymbol{r}^{j(k)} = \left\{r_1^{j(k)}, r_2^{j(k)}, \ldots, r_M^{j(k)}\right\}$;

4. Calculating $U_{s_i}(\boldsymbol{B}^{j(k)}, \boldsymbol{r}^{j(k)}, \boldsymbol{R}^{j(k)})$;

5. Calculating the gradient $\boldsymbol{g}_k = g(\boldsymbol{B}^{j(k)}, \boldsymbol{r}^{j(k)}, \boldsymbol{R}^{j(k)})$;

6. If $\|g_k\| < \varepsilon$, then $\boldsymbol{B}^{j*} \leftarrow \boldsymbol{B}^{j(k)}$, $\boldsymbol{R}^* \leftarrow \boldsymbol{R}^{j(k)}$, and $\boldsymbol{r}^{j*} \leftarrow \boldsymbol{r}^{j(k)}$;

7. Otherwise, let $\boldsymbol{p}_k = -g(\boldsymbol{B}^{j(k)}, \boldsymbol{r}^{j(k)}, \boldsymbol{R}^{j(k)})$, and let $\boldsymbol{B}^{j(k+1)} = \boldsymbol{B}^{j(k)} + \lambda \boldsymbol{p}_k$, $\boldsymbol{R}^{j(k+1)} = \boldsymbol{R}^{j(k)} + \lambda \boldsymbol{p}_k$, and $\boldsymbol{r}^{j(k+1)} = \boldsymbol{r}^{j(k)} + \lambda \boldsymbol{p}_k$;

8. Calculating $U_{s_i}(\boldsymbol{B}^{j(k+1)}, \boldsymbol{r}^{j(k+1)}, \boldsymbol{R}^{j(k+1)}) = U_{s_i}(\boldsymbol{B}^{j(k)} + \lambda \boldsymbol{p}_k, \boldsymbol{r}^{j(k)} + \lambda \boldsymbol{p}_k, \boldsymbol{R}^{j(k)} + \lambda \boldsymbol{p}_k)$;

9. If $\left\|U_{s_i}(\boldsymbol{B}^{j(k+1)}, \boldsymbol{r}^{j(k+1)}, \boldsymbol{R}^{j(k+1)}) - U_{s_i}(\boldsymbol{B}^{j(k)}, \boldsymbol{r}^{j(k)}, \boldsymbol{R}^{j(k)})\right\| < \varepsilon$ or $\max\{\|\boldsymbol{B}^{j(k+1)} - \boldsymbol{B}^{j(k)}\|, \|\boldsymbol{r}^{j(k+1)} - \boldsymbol{r}^{j(k)}\|, \|\boldsymbol{R}^{j(k+1)} - \boldsymbol{R}^{j(k)}\|\} < \varepsilon$;

10. then $\boldsymbol{B}^{j*} \leftarrow \boldsymbol{B}^{j(k+1)}$, $\boldsymbol{R}^* \leftarrow \boldsymbol{R}^{j(k+1)}$, and $\boldsymbol{r}^{j*} \leftarrow \boldsymbol{r}^{j(k+1)}$;

11. otherwise, $k = k + 1$;

12. end if

*# Calculating the optimal strategy of the rest layers #*

13. When $1 < j \leq M$;

*# Loop iteration #*

14. Let $\boldsymbol{B}^{j+1(0)} = \boldsymbol{B}^{j*}$, $\boldsymbol{R}^{j+1(0)} = \boldsymbol{R}^{j*}$, and $\boldsymbol{r}^{j+1(0)} = \boldsymbol{r}^{j*}$, $\forall i \in [1\ X]$ and $\forall j \in [1\ M]$;

15. repeating step 3 to Step 11;

16. $j = j + 1$;

*# Finding the optimal strategy #*

17. Calculating $\boldsymbol{U} = \{U_1(\boldsymbol{B}^{1*}, \boldsymbol{r}^{1*}, \boldsymbol{R}^{1*}), \ldots, U_M(\boldsymbol{B}^{M*}, \boldsymbol{r}^{M*}, \boldsymbol{R}^{M*})\}$;

18. $(\boldsymbol{s}, \boldsymbol{B}, \boldsymbol{r}, \boldsymbol{R}) \leftarrow \arg\min\limits_{\boldsymbol{s}, \boldsymbol{B}, \boldsymbol{r}, \boldsymbol{R}} \boldsymbol{U}$;

The MLi-GD algorithm shown in Table 2 also includes three parts, which are similar to that in Li-GD algorithm. The only difference is that in Li-GD algorithm the variables are $B$ and $r$, which are $B$, $r$, and $R$ in MLi-GD algorithm. Moreover, even we relax $R = \{0,1\}$ to $R \in [0,1]$ in MLi-GD algorithm, we will prove that this relaxation is accuracy in Corollary 7.

For the MLi-GD algorithm, we have the conclusions as follows.

**Corollary 5**. The utility function shown in (43) is differentiable.

*Proof.* For (43), the first term is the same as that shown in (18). In the second term of (43), since the model segmentation strategy and the edge server are not changed, the values of $U_2^{id}$, $U_2^{ie}$, $w_s^i$ and $m_s^i$ keep constant, too. However, due to the routing of intermediate data varies, for reducing inference delay, the allocated bandwidth will change, i.e.,

$B_s^i$ in the second term. Therefore, $U'|_{r_i} = U'_{s_i}|_{r_i}$ and $U'|_R$ is constant. However, due to the $B_s^i$, $U'|_{B_i} \neq U'_{s_i}|_{B_i}$. Fortunately, since the form of $\frac{w_s^i + m_s^i}{B_s^i}$ in the second term is the same as that in the first term, $\frac{w_s^i + m_s^i}{B_s^i}$ is also differentiable. Thus, based on Corollary 1, the utility function in (43) is differentiable. ■

**Corollary 6**. The MLi-GD algorithm is convergent, and the convergence time is $K = \frac{\|x^0 - x^*\|_2^2}{2\eta\epsilon}$, where $\eta$ is the step size and $\eta \leq \frac{1}{L}$, $\epsilon$ is the accuracy.

*Proof.* For (43), since $U'|_{r_i} = U'_{s_i}|_{r_i}$ and $U''|_{r_i} = U''_{s_i}|_{r_i}$, the $U_{s_i}(r_i)$ is continuous, convex and L-Lipschitz. Based on the conclusion in Corollary 5, even $U'|_{B_i} \neq U'_{s_i}|_{B_i}$, the form of $\frac{w_s^i + m_s^i}{B_s^i}$ in the second term is the same as that in the first term, thus, $U_{s_i}(B_i)$ is also continuous, convex and L-Lipschitz. Moreover, $U'|_R$ is constant and $U''|_R = 0$, $U_{s_i}(R_i)$ is continuous, convex and L-Lipschitz. Thus, based on the conclusion in Corollary 2, we can conclude that the MLi-GD algorithm is convergent, and the convergence time is $K = \frac{\|x^0 - x^*\|_2^2}{2\eta\epsilon}$. ■

**Corollary 7**. The approximation ratio of MLi-GD algorithm is $\epsilon$.

*Proof.* For the Li-GD algorithm, the approaximation comes from the accuracy threshold of GD approach. However, in MLi-GD algorithm, except for $\epsilon$, the approximation also comes from $R$, $B$, $r$, and $s$. Recalling that:

$$U = (1 - R) \cdot U_1 + R \cdot U_2 = U_1 + R \cdot (U_2 - U_1) \quad (44)$$

Let $f(R) = U_1 + R \cdot (U_2 - U_1)$, then, if $U_2 > U_1$, $f'(R) > 0$; if $U_2 < U_1$, $f'(R) < 0$. Considering that $R \in [0\ 1]$, if $U_2 > U_1$, $f(R)$ is monotone increasing, thus, only $R = 0$, the $f(R)$ can get the minimum value; if $U_2 < U_1$, $f(R)$ is monotone decreasing, thus, only $R = 1$, the $f(R)$ can get the minimum value. Therefore, the approximation ratio for relaxing $R \in \{0, 1\}$ to $R \in [0\ 1]$ is 1.

Additionally, since the model segmentation strategy in $U_2$ has been decided, the resource allocation strategies of $B$ and $r$ are also decided. Therefore, for different model segmentation and resource allocation strategies in $U_1$, only $R$ changes. Assuming that $U'_1(B, r, s)$ is the optimal value of $U_1$ in (44) and $U_1^*(B^*, r^*, s^*)$ is the optimal value of (18), then let:

$$U' = (1 - R') \cdot U'_1 + R' \cdot U_2 \quad (45)$$
$$U^* = (1 - R^*) \cdot U_1^* + R^* \cdot U_2 \quad (46)$$

Assume that $U'_1 < U_2$ and $U_1^* < U_2$, since $U'_1 > U_1^*$, $R^* < R'$ holds. Then,

$$\begin{aligned} U^* - U' &= (1 - R^*) \cdot U_1^* - (1 - R') \cdot U'_1 + (R^* - R') \cdot U_2 \\ &< (1 - R^*) \cdot U'_1 - (1 - R') \cdot U'_1 + (R^* - R') \cdot U_2 \\ &= (R' - R^*) \cdot U'_1 + (R^* - R') \cdot U_2 \\ &= (R' - R^*) \cdot (U'_1 - U_2) < 0 \end{aligned} \quad (47)$$

Therefore, once the optimal $U'_1(B, r, s)$ is calculated based on (43), it equals to $U_1^*(B^*, r^*, s^*)$. This means that the approximation ratio regards to $B$, $r$, and $s$ is 1, too.

Then, considering that the accuracy of GD algorithm is $\epsilon$, we can conclude that the final approximation ratio of MLi-GD algorithm is $\epsilon$. ■

**Corollary 8**. The complexity of MLi-GD algorithm is the same as that of the Li-GD algorithm.

*Proof.* For the algorithm which takes the mobility into account, on the one hand, $U'|_{r_i} = U'_{s_i}|_{r_i}$ and $U'|_R$ is constant;

on the other hand, the complexity of $U'|_{B_i}$ is much higher than that of $U'|_{r_i}$ and $U'|_R$. Even $U'|_{B_i} \neq U'_{S_i}|_{B_i}$, the complexity of $\frac{w_S^i + m_S^i}{B_S^i}\Big|_{B_S^i}$ in the second term is the same as that in the first. Moreover, the complexity of $U'|_{B_i}$ is mainly determined by $p_i \frac{w_{S_i}}{B_i \log_2\left(1 + \frac{p_i \alpha_i^\kappa g_i^\kappa}{B_i N_0}\right)}$, which is not affected by $\frac{w_S^i + m_S^i}{B_S^i}$ in the second term of (43). Additionally, the complexity of $U'|_{B_i}$ is $(x^3 \ln^2(x))$. Then, the complexity of MLi-GD algorithm in worst cause is $O(XK_{MLi}^* Mx^3 \ln^2(x))$, where $K^*$ is the maximum convergence time during $M$ layers and $K_{MLi}^* = \max\{K_{MLi}^1, K_{MLi}^2, \dots, K_{MLi}^M\}$. The average complexity of the MLi-GD algorithm is $O(X\overline{K}_{MLi} Mx^3 \ln^2(x))$, where $\overline{K}_{MLi} = \frac{\sum_{i=1}^M K_{MLi}^i}{M}$. ∎

Note that the complexity that calculated in Li-GD algorithm and MLi-GD algorithm are samiliar. Even the number of varibables is larger in MLi-GD algorithm than that in Li-GD algorithm, considering that the complexity of $U|_{B_i}$ is much higher than $U|_{r_i}$ and $U|_{R_i}$, the $R_i$ in MLi-GD algorithm will not increase the complexity of GD process. Moreover, in MLi-GD algorithm, for finding the optimal model segmentation point, the delay for transmitting the inference task back to the original edge server should be calculated. However, in this process, since the model segmentation strategy is known, the complexity mainly come from the shortest transmission time path calculation whose complexity is $O(n^2)$ (where $n$ is the number of related APs), which is much smaller than the complexity of GD process. Therefore, the complexity of Li-GD algorithm and MLi-GD algorithm is similar.

# 6 PERFORMANCE EVALUATION

## 6.1 Experimental Setup

Dataset. We use CIFAR-10 dataset in this paper. The CIFAR-10 dataset consists of 60000 $32 \times 32$ RGB images in 10 classes (from 0 to 9), with 50000 training images and 1000 test images per class.

DNN benchmarks. There are many DNN models with different topologies have been proposed recently. For instance, NiN, tiny YOLOv2, VGG16, etc., are the well-known chain topology models; AlexNet, ResNet-18, etc., are the well-known DAG topology models. However, in this paper, we mainly evaluate the performance of the proposed algorithms on chain topology models, i.e., NiN (9 layers), YOLOv2 (17 layers) and VGG16 (24 layers).

Evaluation benchmarks. We compare the proposed algorithms against Device-Only (i.e., executing the entire DNN on the device), Edge-Only (i.e., executing the entire DNN on the edge), Neurosurgeon, and DNN surgeon. However, the DNN surgeon can operate on both chain topology models and DAG topology models. In this paper, since we mainly focus on chain topology, we only implement DNN surgeon on chain topology models (i.e., NiN, YOLOv2, and VGG16).

## 6.2 Performance evaluation without user mobility

In this section, we compare the performance of MCSA with Device-Only, Edge-Only, Neurosurgeon, and DNN surgeon under the scenario in which the users are not mobile. In this section, we use the Device-Only method as the baseline, i.e., the performance is normalized to the Device-Only method.

Compare the MCSA with Device-Only and Edge-Only. The performance of MCSA has been compared with Device-Only and Edge-Only strategies. The performance of latency speedup, energy consumption reduction, and resource renting cost is presented in Fig.3, Fig.4, and Fig.5, respectively. In this section, we use the Device-Only method as the baseline, i.e., the performance is normalized to the Device-Only method.

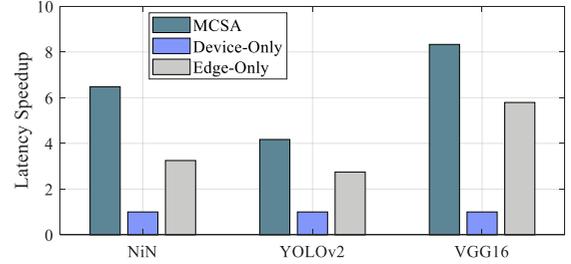

Fig.3. Latency speedup for different DNN models

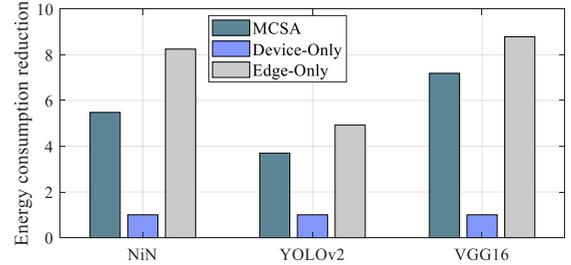

Fig.4. Energy consumption reduction for different DNN models

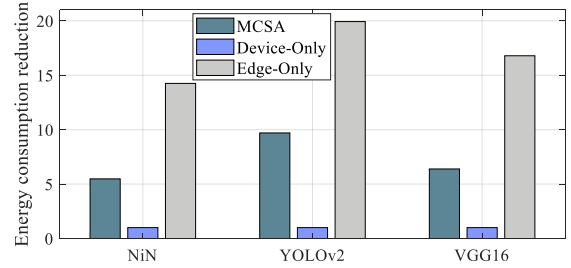

Fig.5. Resource renting cost for different DNN models

From Fig.3, we can find that comparing with Device-Only approach, both the MCSA and Edge-Only approach can reduce the inference latency. For instance, the latency speedup of MCSA is 4.08 to 8.2 times higher than the Device-Only approach. This is because the Device-Only approach, the whole inference task is executed on device. However, the computing capability of device is lower than the edge server. Moreover, the performance of MCSA approach is also better than that of the Edge-Only approach. Since even the computing capability of edge server is better than the device, besides considering the fact that the size of raw data is always huge, which makes the data transmission delay high. Therefore, the performance of MCSA is better than the Edge-Only approach.

The performance of energy consumption reduction is presented in Fig.4. We can find that both the MCSA and Edge-Only approaches can reduce the energy consumption of devices. For instance, the energy consumption reduction of MCSA is 3.8 to 7.1 times higher than Device-Only approach. Moreover, the energy consumption reduction much more in Edge-Only approach than that in MCSA. Since in the Edge-Only approach, the whole inference task is executed on edge, there is no task on device; for the MCSA, only part of the inference task is executed on device; for the Device-Only approach, the whole inference task is executed on device.

The performance of resource renting cost is presented in Fig.5. We can conclude that the resource renting cost in Edge-Only approach is the highest and the MCSA takes the second place. For instance, the resource renting cost in MCSA is 5.5 to 9.7 times higher than the Device-Only approach. This is easy to understand. Because for the Edge-Only approach, on the one hand, the whole inference task needs to be calculated on edge, thus, the rented computing resource is the largest; on the other hand, for the Edge-Only approach, the size of the transmitted to edge server is the biggest, thus, the size of the rented communication resource is the largest. Therefore, the resource renting cost in Edge-Only is the biggest. Moreover, since the MCSA takes the resource renting cost during finding the optimal strategy, the resource renting cost is much lower than that in Edge-Only approach.

Comparing MCSA with Neurosurgeon and DNN surgeon. The performance of MCSA is compared with Neurosurgeon and DNN surgeon. The performance of latency speedup, energy consumption reduction, and resource renting cost is presented in Fig.6, Fig.7, and Fig.8, respectively. In this section, we use the Neurosurgeon as the baseline, i.e., the performance is normalized to the Neurosurgeon method.

From Fig.6, we can find that the latency speedup in MCSA and DNN surgeon is similar but a little better than that in Neurosurgeon. For instance, the latency speedup in MCSA is 0.89 to 0.92 times smaller than that in the Neurosurgeon. And the performance of DNN surgeon is also similar but a little better than MCSA. This is because in MCSA, we take the energy consumption and the resource renting into accout to find the optimal tradeoff between latency, energy consumption, and resource renting, which is not considered in Neruosurgeon and in DNN surgeon, the amount of allocated computing resource is limited. Therefore, the computiong resource and communication resource that allocated in MCSA is less than the other two algorithms, which causes reduction of latency speedup.

The performance of energy consumption reduction is presented in Fig.7. We can find that the energy consumption reduction in MCSA is much better than that in Neurosurgeon and DNN surgeon, and the energy consumption reduction in Neurosurgeon and DNN surgeon is similar. For instance, the energy consumption reduction in MSCA is 1.8 to 2.48 times larger than Neurosurgeon. The energy consumption reduction in DNN surgeon is similar to that in Neurosurgeon but a little worse. Since in both Neurosurgeon and DNN surgeon, they do not consider the energy consumption of mobile device. Additionally, DNN surgeon considers the resource limitation of edge server; thus, comparing with Neurosurgeon, the energy consumption of mobile device in DNN surgeon will be a little high.

The performance of resource renting cost is presented in Fig.8. We can conclude that the performance of resource renting cost in MCSA is much better than that in Neurosurgeon and DNN surgeon. For instance, the resource renting cost in MCSA is 0.76 to 0.81 times lower than Neruosurgeon. The performance of DNN surgeon is better than that of Neurosurgeon but worse than that of MCSA. This is because in MCSA, we minimize the resource renting cost. Additionally, DNN surgeon considers the resource limitation of edge server; thus, comparing with Neurosurgeon, the allocated computing resource to user will be smaller than that in Neurosurgeon.

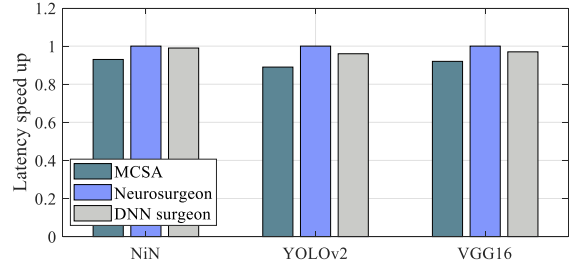

Fig.6. Latency speedup for different DNN models

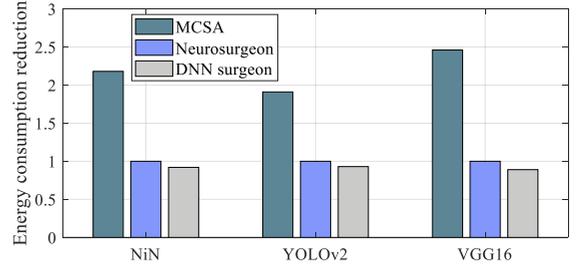

Fig.7. Energy consumption reduction for different DNN models

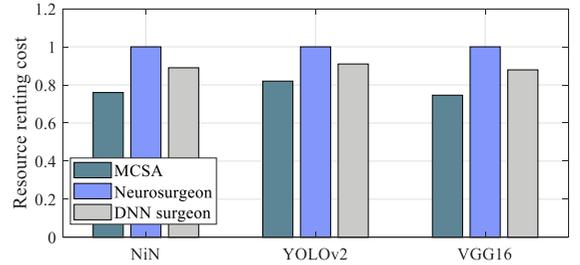

Fig.8. Resource renting cost for different DNN models

## 6.3 Performance evaluation with user mobility

In this section, we compare the performance of MCSA with Device-Only, Edge-Only, Neurosurgeon, and DNN surgeon under the scenario in which the users are mobile.

Comparing MCSA with Device-Only and Edge-Only. The performance of MCSA has been compared with Device-Only and Edge-Only strategies. The performance of latency speedup, energy consumption reduction, and resource renting cost is presented in Fig.9, Fig.10, and Fig.11, respectively. In this section, we use the Device-Only method as the baseline, i.e., the performance is normalized to the Device-Only method.

From Fig.9, we can find that comparing with Device-Only approach, both the MCSA and Edge-Only approach can reduce the inference latency. For instance, the latency speedup of MCSA is 3.9 to 7.2 times higher than the Device-Only approach. The reasons are the same as that in Fig.3. however, comparing with Fig.3, the latency speedup in MCSA and DNN surgeon is reduced in Fig.9. This is because when the device is mobile, for the DNN surgeon, the intermediate data transmission routing increases, which causes the reduction of latency speed up. For the MCSA, since the mobile user needs to recalculate optimal strategy, the latency increases, too.

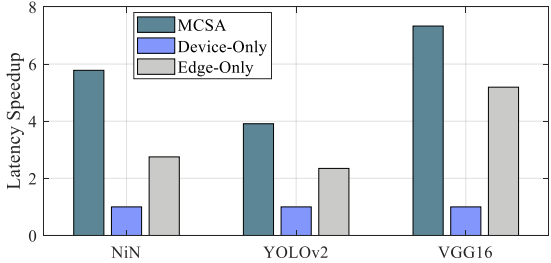

Fig.9. Latency speedup for different DNN models

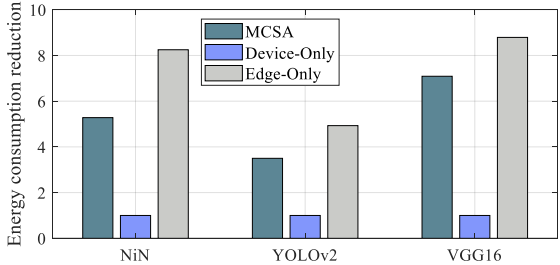

Fig.10. Energy consumption reduction for different DNN models

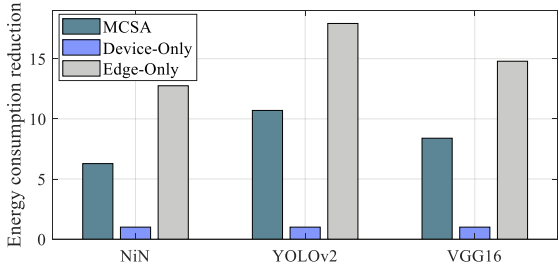

Fig.11. Resource renting cost for different DNN models

The performance of energy consumption reduction is presented in Fig.10. We can find that both the MCSA and Edge-Only approaches can reduce the energy consumption of devices. For instance, the energy consumption reduction of MCSA is 3.4 to 6.9 times higher than Device-Only approach. Moreover, the energy consumption reduction in Edge-Only approach is much more than that in MCSA. The reason is the same as that in Fig.4. Moreover, comparing with the results in Fig.4, only the energy consumption reduction in MCSA is decreased, the energy consumption reduction in Edge-Only approach is not affected by the user mobility. Since in the Edge-Only approach, the whole inference task is offloaded to the edge server, the energy consumption is nothing to do with the device; only the intermediate data transmission delay is affected.

The performance of resource renting cost is presented in Fig.11. We can conclude that the resource renting cost in Edge-Only approach is the highest and the MCSA takes the second place. For instance, the resource renting cost in MCSA is 6.3 to 10.7 times higher than the Device-Only approach. The reason is the same as that in Fig.5. However, for the resource renting cost, considering the user mobility and the heterogeneity of edge servers, the performance of MCSA and Edge-Only approaches change with different edge servers. For instance, the results in Fig.5 and Fig.11 are different.

Comparing MCSA with Neurosurgeon and DNN surgeon. The performance of MCSA is compared with Neurosurgeon and DNN surgeon. The performance of latency speedup, energy consumption reduction, and resource renting cost is presented in Fig.12, Fig.13, and Fig.14, respectively. In this section, we use the Neurosurgeon as the baseline, i.e., the performance is normalized to the Neurosurgeon method.

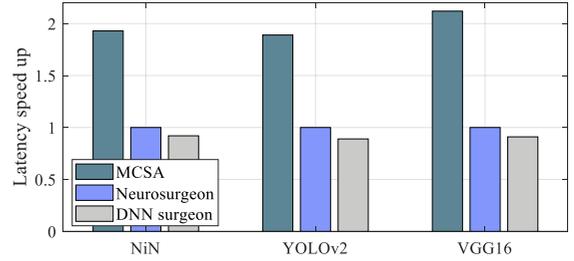

Fig.12. Latency speedup for different DNN models

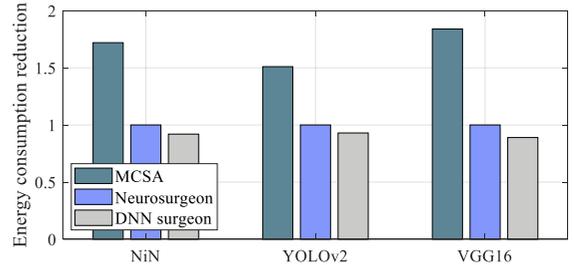

Fig.13. Energy consumption reduction for different DNN models

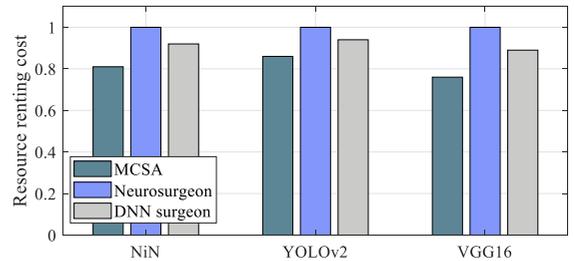

Fig.14. Resource renting cost for different DNN models

From Fig.12, we can find that the latency speedup in Neurosurgeon and DNN surgeon is similar but worse than that in MCSA. For instance, the latency speedup in MCSA is 1.9 to 2.2 times higher than the Neurosurgeon. The performance of DNN surgeon is similar to but a little worse than Neurosrugeon. This is different with the results in Fig.6. The reason is that in MCSA, we take the user mobility into account to find the optimal model segemntation and resource allocation strategy fo minimize the inference latency. However, since the DNN surgeon assumes that the amount of computing resource is limited in edge server, the latency in DNN surgeon is a little worse than that in Neurosurgeon.

The performance of energy consumption reduction is presented in Fig.13. We can find that the energy consumption reduction in MCSA is much better than that in Neurosurgeon and DNN surgeon, and the energy consumption reduction in Neurosurgeon and DNN surgeon is similar. For instance, the energy consumption reduction in MSCA is 1.5 to 1.8 times larger than Neurosurgeon. The energy consumption reduction in DNN surgeon is similar to Neurosurgeon but a little worse. This is because in both Neurosurgeon and DNN surgeon, they do not consider the energy consumption and mobility of device. Additionally, DNN surgeon takes the resource limitation of edge server into account; thus, comparing with Neurosurgeon, the energy consumption of mobile device in DNN surgeon will be a little high.

The performance of resource renting cost is presented in Fig.14. We can conclude that the performance of resource renting cost in MCSA is much better than that in Neurosurgeon and DNN surgeon. For instance, the resource renting cost in MCSA is 0.78 to 0.85 times lower than Neruosurgeon. The performance of DNN surgeon is better than that of Neurosurgeon but worse than that of MCSA. This is because in MCSA, we minimize the resource renting cost with user mobility. Additionally, DNN surgeon considers the resource limitation of edge server; thus, comparing with Neurosurgeon, the allocated computing resource to user will be smaller than that in Neurosurgeon.

## 6.4 Performance evaluation under different network condition

In this section, we compare the performance of MCSA with Device-Only, Edge-Only, Neurosurgeon, and DNN surgeon under different network conditions, including different number of hops for intermediate data transmission and inference task computing load. The results are presented in Fig.15 and Fig.16, respectively. In this section, we use the Device-Only method as the baseline, i.e., the performance is normalized to the Device-Only method.

The latency of MCSA, Device-Only, Edge-Only, Neurosurgeon, and DNN surgeon under different number of hops for intermediate data transmission is presented in Fig.15. We can conclude that with the increasing of the number of hops, i.e., the distance between the mobile devices to its original edge server, the latency in Edge-Only, Neurosurgeon, and DNN surgeon rises. Since in Device-Only approach, the whole inference modle is in the mobile device, its latency performance is not affected by the number of hops. For MCSA, with the increasing of the number of hops, its latency is relatively stable compeared with Edge-Only, Neurosurgeon, and DNN surgeon. The latency performance of MCSA is the best, and with the increasing of the number of hops, the advantage of MCSA becomes more and more obviously. For instance, when $N = 2$, the latency speedup of MCSA, Edge-Only, Neurosurgeon, and DNN surgeon is 8.2 times, 6.17 times, 7.95 times, 7.8 times, respectively; when $N = 10$, these values become 8.24 times, 1.86 times, 3.87 times, and 3.66 tiems, respectively. This is because, except for the MCSA, the Edge-Only, Neurosurgeon, and DNN surgeon approaches do not consider the user mobility, thus, when the users move far away from the original edge server, the intermediate data transmission gains. The MCSA considers the user mobility, it can stable the performance of inference latency.

The performance of lantency speedup under different inference task computing load is presented in Fig.16. We can find that with the increasing of the computing load, the performance of latency speedup of all the algorithms except for the Device-Only approach. In the Device-Only approach, since the whole inference model is executed on device, when the number of inference rounds grow, the average delay of each round is not affected. In other algorithms, when the number of inference rounds enhance, the load in the communication link will increase, i.e., the communication resource becomes shortage, which will rise the intermediate data transmission delay. Therefore, all the latency increases in MCSA, Edge-Only, Neurosurgeon, and DNN surgeon algorithm. However, considering that in MCSA, we take the communication resource allocation into account, the delay increasing in MCSA is smaller than the other algorithms. Moreover, since the size of intermediate data in Edge-Only approach is much larger than that of the Neurosurgeon and DNN surgeon, its inference delay is higher than that of the Neurosurgeon and DNN surgeon.

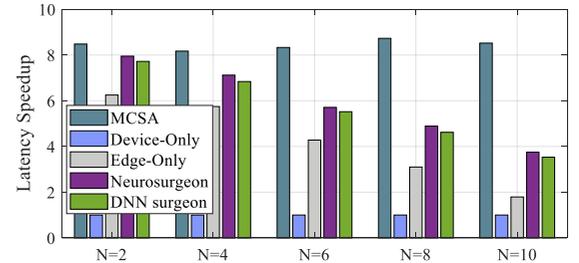

Fig.15. Latency speedup under different number of hops

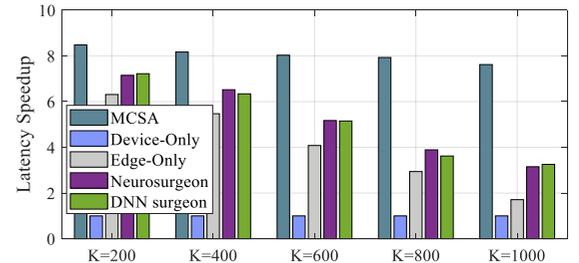

Fig.16. Latency speedup under different computing load

# 7. CONCLUSION

In this paper, for addressing the disadvantages of model segmentation and resource allocation in previous works, we propose the MCSA. In the scenario without user mobility, when the mobile user has a large model inference task needs to be calculated, it will take the energy consumption of mobile user, the communication and computing resource renting cost, and the inference delay into account to find the optimal model segmentation strategy and resource allocation strategy. Since the minimum delay, minimum cost, and minimum energy consumption cannot be satisfied simultaneously, we use the GD algorithm to find the optimal tradeoff between them. Moreover, we propose the Li-GD algorithm to reduce the complexity of GD algorithm that caused by the discrete of model segmentation. In the scenario which considering user mobility, since there are multiple alternative strategies, for finding the optimal solution between these strategies, we introduce the variable $R$ into the objective function to reprsent the alternative strategies. Then, we propose the MLi-GD algorithm to calculate the optimal model segmentation, resource allocation, and strategy selection under user mobility. Finally, we investigate the properties of the proposed algorithms, including convergence, complexity, and approximation ratio. The experimental results demonstrate the effectiveness of the proposed algorithms.

## ACKNOWLEDGMENT

This work was supported in part by a grant from NSFC Grant no. 62101159, NSF of Shandong Grant no. ZR2021MF055, and also the Research Grants Council of Hong Kong under the Areas of Excellence scheme grant AoE/E-601/22-R.